\documentclass[prb,aps,amssymb,shownopacs,twocolumn,10pt]{revtex4-1}
\usepackage{amsmath}
\usepackage{amssymb}
\usepackage{amsthm}
\usepackage{amsfonts}
\usepackage{listings}
\usepackage{easybmat}
\usepackage{enumerate}
\usepackage{latexsym}
\usepackage{epstopdf}
\usepackage{psfrag}
\usepackage{color}
\usepackage{bm}
\usepackage[pdftex]{graphicx}
\usepackage{subfigure}
\usepackage{hyperref}
\usepackage{ulem} 
\newcommand{\beq}{\begin{equation}}
\newcommand{\eneq}{\end{equation}}
\newcommand{\bs}[1]{\boldsymbol{#1}}

\newcommand{\braket}[2]{\left\langle #1 | #2 \right\rangle}

\newcommand{\bra}[1]{\left\langle#1\right|}
\newcommand{\ket}[1]{\left|#1\right\rangle}

\newcommand{\rtilde}{\tilde{\rho}}

\newcommand{\pref}[1]{(\ref{#1})}

\def\be{\begin{equation}}
\def\ee{\end{equation}}
\def\ba{\begin{eqnarray}}
\def\ea{\end{eqnarray}}

\def\mbf{\mathbf}

\def\a{\alpha}

\def\g{\gamma}

\def\D{\Delta}

\def\n{\nu}

\input{epsf}

\begin{document}

\tolerance 10000
\newcommand{\vk}{{\bf k}}
\title{
The single-mode approximation for fractional Chern insulators and the fractional quantum Hall effect on the torus
}

\author{C\'ecile Repellin$^1$}
\author{Titus Neupert$^2$} 
\author{Zlatko Papi\'c$^{3,4}$}
\author{Nicolas Regnault$^{5,1}$}

\affiliation{$^1$ Laboratoire Pierre Aigrain, ENS-CNRS UMR 8551, Universit\'es P. et M. Curie and Paris-Diderot, 24, rue Lhomond, 75231 Paris Cedex 05, France\\
$^2$ Princeton Center for Theoretical Science, Princeton University, Princeton, NJ 08544, USA\\
$^3$ Perimeter Institute for Theoretical Physics, Waterloo, ON N2L 2Y5, Canada\\ 
$^4$ Institute for Quantum Computing, Waterloo, ON N2L 3G1, Canada\\
$^5$ Department of Physics, Princeton University, Princeton, NJ 08544, USA} 

\begin{abstract}
We analyze the collective magneto-roton excitations of bosonic Laughlin $\nu=1/2$
fractional quantum Hall (FQH) states on the torus and of their analog on the lattice, the fractional Chern insulators (FCIs).
We show that, by applying the appropriate mapping of momentum quantum numbers between the two systems,
the magneto-roton mode can be identified in FCIs and that it contains the same number of states as in the FQH case. 
Further, we numerically test the single mode approximation to the magneto-roton mode for both the FQH and FCI case. 
This proves particularly challenging for the FCI, because its eigenstates have a lower translational symmetry than the FQH states. In spite of this, we construct the FCI single-mode approximation such that it carries the same momenta as the FQH states, allowing for a direct comparison between the two systems.
We show that the single-mode approximation captures well a dispersive subset of the magneto-roton excitations both for the FQH and the FCI case. We find remarkable quantitative agreement between the two systems. For example, the many-body excitation gap extrapolates to almost the same value in the thermodynamic limit.
\end{abstract}

\date{\today}

\maketitle

\section{Introduction}
It is a celebrated property of topologically ordered gapped ground states of quantum matter 
that they contain the \emph{universal} information about their topological excitations. 
For example, the statistics of quasiparticles in Abelian fractional quantum Hall (FQH) states 
can be inferred from their entanglement spectrum~\cite{li-08prl010504, sterdyniak-PhysRevLett.106.100405} or their response to modular transformations~\cite{Zhang-PhysRevB.85.235151, Moradi-2014arXiv1401.0518M}. 
In the limit where the energy gap is infinite, this information 
completely determines the universal physics inscribed in the state.
However, this universal data does not contain information about the dynamics, i.e, energetics, of excitations above the state. 
As such, it is incapable of explaining whether and why a featureless topological ground state is incompressible 
or what the nature of competing states is.
The answers to these questions
can be inferred from the study of the \emph{nonuniversal} dynamics of its collective excitations.

Fractional quantum Hall states are a class of topologically ordered states, for which both universal and dynamical properties of the collective excitations are well understood. In a seminal work, Girvin, MacDonald, and Platzman (GMP) unraveled that FQH states possess a neutral collective
excitation, the magneto-roton mode, which, at long wavelenghts, is well described by a density wave on top of the featureless ground state 
using the single mode approximation (SMA).~\cite{GMP-PhysRevLett.54.581, GMP-PhysRevB.33.2481} 
Within the SMA, it is possible to demonstrate the existence of a spectral gap above the FQH ground state and study the transition to the Wigner crystal which occurs via softening of the magneto-roton mode. These properties are intimately related to the fact that the density operators, when projected in a single Landau level, do not commute, but rather furnish what is now called the GMP algebra.
The work by GMP was complemented via the construction of explicit wavefunctions for the magneto-roton mode on the sphere for both Abelian and non-Abelian FQH states using a composite fermion approach~\cite{Sreejith-PhysRevLett.107.086806, Rodriguez-PhysRevB.85.035128} and a Jack polynomial approach~\cite{Bernevig-PhysRevLett.100.246802, Bernevig-PhysRevB.77.184502, Bernevig-PhysRevLett.102.066802, Yang-PhysRevLett.108.256807}.

Recently, it was shown that there exist analogues of FQH states for repulsively interacting electrons in lattice models with appropriate topological energy bands~\cite{neupert-PhysRevLett.106.236804,sheng-natcommun.2.389,regnault-PhysRevX.1.021014} (see also Refs.~\cite{Bergholtz-2013IJMPB..2730017B,Parameswaran-2013CRPhy..14..816P} and references therein). In contrast to the FQH effect in Landau levels, no externally applied magnetic field is required in the lattice models. By studying their entanglement spectrum~\cite{regnault-PhysRevX.1.021014}, and via modular~\cite{Cincio-PhysRevLett.110.067208} as well as adiabatic~\cite{Wu-PhysRevB.86.085129,Liu-PhysRevB.87.035306, Scaffidi-PhysRevLett.109.246805, Wu-PhysRevB.86.165129} transformations, it has been established that these so-called fractional Chern insulators (FCI) have excitations with the same topological properties.
However, FCI and FQH states differ in other respects. Importantly, the lattice Hamiltonians with FCI ground states lack the center of mass translational symmetry of the Landau level problem. As a consequence, the GMP algebra of density operators holds only approximately in the limit of long wavelengths \cite{Parameswaran-2012PhRvB..85x1308P,goerbig-2012epjb,Neupert-PhysRevB.86.035125,estienne-PhysRevB.86.241104,Lee-PhysRevB.88.035101}.  
Given these similarities and differences, it is instructive to study the fate of the dynamical collective excitations of FHQ states in FCIs. This serves as one main objective for our work.

We focus on the simplest FQH state, namely the bosonic Laughlin state at filling $\nu=1/2$, and its FCI cousin.
The FCI, defined as a lattice model with periodic boundary conditions, should naturally be compared to the continuum FQH state on the torus geometry. 
To the best of our knowledge, even for this simplest FQH state, no numerical study of the SMA to the magneto-roton mode has been performed on the torus. Hence, the other main objective for the current study is to establish how well the SMA approximates excitations above the FQH Laughlin states on the torus. The results will serve as a reference that allows us to identify the magneto-roton excitations above the FCI ground state. 

To achieve this, we first have to find the correct interpretation of the SMA for the FQH effect in the torus geometry. Using the density wave excitations proposed by GMP, one can build a factor of $\nu^{-2}$ more variational SMA states than the magneto-roton mode contains.
We give a prescription how the SMA states that best approximate the magneto-roton mode should be selected from this manifold of variational states. Equipped with this selection criterion, we find the magneto-roton dispersion of the FQH state well captured by the SMA at long wavelengths. For shorter wavelength, however, the magneto-roton dispersion flattens out, while the SMA dispersion merges with the continuum of multiple quasiparticle-quasihole pair excitations. We show that this behavior is not improved if the space of variational states is enlarged to all SMA states. It is thus not a shortcoming of our selection criterion, but simply reflects the fact that the SMA is not a good approximation to the magneto-roton mode at large momenta. We also give a finite size extrapolation of the gap above the $\nu=1/2$ bosonic Laughlin state, determined by the minimum of the magneto-roton dispersion, to the thermodynamic limit.

Turning to FCIs, we indeed observe a neutral collective mode separated from the quasi-continuum of excited states. However, not all lattice models for FCIs expose this collective mode. We find the mode clearly separated in models which show a smaller finite-size splitting of their quasi-degenerate topological ground states, such as the kagome lattice model~\cite{tang-PhysRevLett.106.236802} and the ruby lattice model~\cite{hu-PhysRevB.84.155116}.
We identify this collective mode as the analogue of the magneto-roton mode of the FQH states with the aid of two complementary pieces of evidence.
First, using the FQH-to-FCI mapping introduced in Ref.~\cite{Bernevig-2012PhysRevB.85.075128}, we compare the number of states belonging to the magneto-roton mode in the FQH and FCI cases. We show that the number of states per momentum sector in the FCI spectrum can be deduced from the FQH counting, provided the FQH torus has the same angle as the torus defined by the periodic boundary conditions of the FCI lattice. 
Second, we develop a systematic procedure to construct the SMA for the magneto-roton mode of the FCI. Due to the absence of the magnetic translations,  building an SMA for FCI is more challenging. Similarly to the FQH case, the FCI density operator allows to build a larger space of variational states for the magneto-roton mode, provided one allows for the momenta of the density wave excitations to lie outside of the first Brillouin zone. 
We give a criterion that determines which momenta of the full Brillouin zone are relevant for the SMA. We then show numerically that these SMA states provide an accurate description of the dispersive subset of magneto-roton states in the FCI. As may be expected from the FQH case, the flat part of the magneto-roton curve cannot be captured by the SMA. Finally, we quantitatively compare the quality of the SMA description of the magneto-roton mode in the FQH and FCI cases. We show that the dispersive branch of the magneto-roton mode above the Laughlin $\nu = 1/2$ state of the ruby lattice model is equally well described by the SMA as that of the FQH.

The paper is structured as follows.
We start off in Sec.~\ref{ref:magnetoroton} 
by presenting the exact diagonalization 
spectra for the FQH and FCI cases, as well as the 
folding of the FQH Brillouin zone to the FCI Brillouin zone
that allows for their direct comparison.
Subsequently, in Sec.~\ref{ref: SMA}, 
we develop the analytical formalism for the SMA for both the FQH effect on the torus and for the FCI.
Section~\ref{ref: SMA data} contains the numerical results of the SMA that test our analytical formalism. 
We conclude our results in Sec~\ref{ref: Conclusion}.

\section{Numerical evidence for the magneto-roton mode}
\label{ref:magnetoroton}
\subsection{Fractional quantum Hall system on the torus}
\label{ref:magnetorotonFQH}
\begin{figure}
\includegraphics[width = 0.8\linewidth]{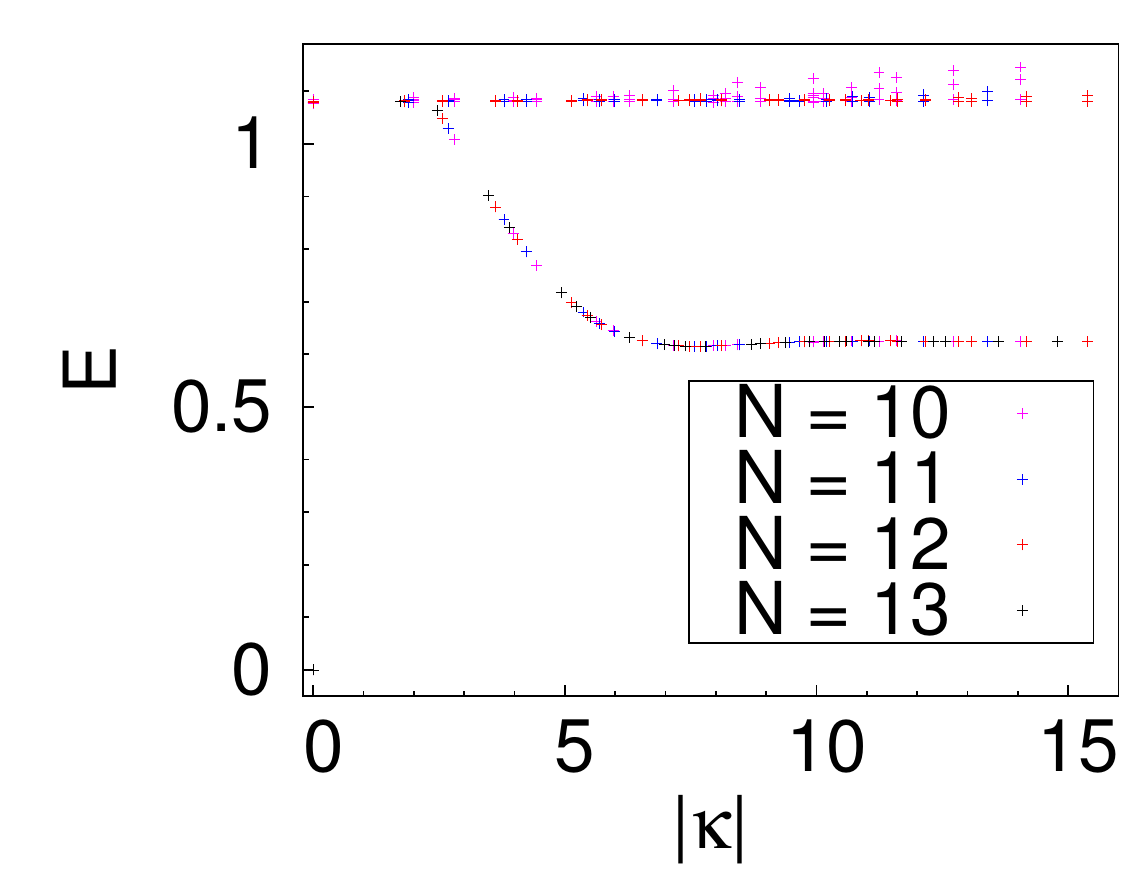}
\caption{Low energy spectrum of the FQH on the torus for up to $N = 13$ bosons and $N_{\phi} = 2\times N$ flux quanta, as a function of the modulus of the momentum $\bs \kappa$, where $\bs \kappa$ is defined Eq.~\eqref{eq: K-rescaling for data collapse}.}
\label{fig:magnetoroton}
\end{figure}
We consider a system of $N$ bosons on a torus pierced by $N_{\phi}$ flux quanta. The torus is spanned by the vectors $L_x \bs{e}_x$ and $L_y \bs{e}_y$, where $\bs{e}_x$ and $\bs{e}_y$ are two unit vectors. Unless otherwise specified, we consider a square torus, i.e. the aspect ratio $L_x/L_y$ equals $1$, and the twisting angle is $\theta = \pi/2$ ($\cos\theta \equiv \bs{e}_x\cdot \bs{e}_y$). While all our numerical results are for the fixed filling $\n = N/N_{\phi} = 1/2$ of the lowest Landau level (LLL), we will give the analytical expressions in this section for a generic bosonic filling $\n = 1/m$, where $m$ is an even integer. The bosons interact via the repulsive delta interaction, which is the model Hamiltonian for which the bosonic $\n = 1/2$ Laughlin state is the densest zero-energy state. In the pseudo-potential language\cite{Haldane-PhysRevLett.51.605}, it means that we consider only the $V_0$ pseudo-potential.

Translation operators on the torus can be factorized into the product of a center of mass (CM) and a relative translation. The CM translation operator along the $y$ axis and the relative translation operator along the $x$ axis commute with each other and with the Hamiltonian. The eigenstates of the Hamiltonian thus carry the corresponding momentum quantum numbers $\bs{k}$ that belong to the FQH Brillouin zone
\be
\begin{split}
&\mathrm{BZ}_{\mathrm{FQH}}
\equiv\left\{
\bs{k}=\frac{2\pi}{L_x}\mathsf{k}_x \bs{\tilde{e}}_x + \frac{2\pi}{L_y}\mathsf{k}_y \bs{\tilde{e}}_y
\right|
\\
&\left.
\mathsf{k}_x=0,\cdots, \mathrm{GCD}(N,N_\phi)-1;
\
\mathsf{k}_y=0,\cdots, N_\phi-1
\right\}.
\end{split}
\ee
where GCD stands for the greatest common divisor, and $\bs{\tilde{e}}_x, \bs{\tilde{e}}_y$ are two unit vectors in the reciprocal lattice, such that $\bs{e}_i \cdot \bs{\tilde{e}}_j = \delta_{i,j}$. In the following, we shall only consider cases where $N_\phi=mN$, so that $\mathrm{BZ}_{\mathrm{FQH}}$ consist of $N\times N_\phi$ points.   
To observe the magneto-roton mode, the spectrum should be plotted as a function of $|\bs{k}|$, where
$\bs{k}$ takes values in the reduced Brillouin zone~\cite{Haldane85-PhysRevLett.55.2095} of size $N \times N$
\be
\begin{split}
\mathrm{BZ}^{\mathrm{red}}_{\mathrm{FQH}}
\equiv&\left\{
\bs{k}=\frac{2\pi}{L_x}\mathsf{k}_x \bs{\tilde{e}}_x + \frac{2\pi}{L_y}\mathsf{k}_y \bs{\tilde{e}}_y
\right|
\\
&\left.
\mathsf{k}_x=0,\cdots, N-1;
\
\mathsf{k}_y=0,\cdots, N-1
\right\}.
\end{split}
\ee
While the $m$ topologically degenerate ground states appear at different momenta $\bs{\mathsf{K}}_\alpha\equiv(0,\alpha N)\in\mathrm{BZ}_{\mathrm{FQH}}$, with $\alpha=0,\cdots, m - 1$, they all map to momentum $\bs{\mathsf{k}}=0$ in $\mathrm{BZ}^{\mathrm{red}}_{\mathrm{FQH}}$. This way, all their magneto-roton dispersions coincide at the same momenta.
Collapsing data for various system sizes, Fig.~\ref{fig:magnetoroton} clearly exposes the magneto-roton mode as a excitation mode above the ground state and below the continuum of higher energy excitations. 
In order to obtain the data collapse, all momenta $\bs{k}\in\mathrm{BZ}^{\mathrm{red}}_{\mathrm{FQH}}$ have to be rescaled by a factor $1/\sqrt{N\,\sin(\theta)/(L_xL_y)}$ to yield a dimensionless momentum $\bs{\kappa}$ that is defined in a Brillouin zone of area $N\,(2\pi)^2$. The data is then plotted as a function of $|\bs{\kappa}|$, i.e.,
\be
|\bs{\kappa}|
=\frac{2\pi}{\sqrt{N\,\sin\theta}}
\sqrt{r^{-1}\mathsf{k}^2_x+r\mathsf{k}^2_y-2\cos\theta\,\mathsf{k}_x\mathsf{k}_y},
\label{eq: K-rescaling for data collapse}
\ee
where $r=L_x/L_y$ is the aspect ratio.
We observe that the magneto-roton dispersion does not show a magneto-roton minimum, but rather flattens out for momenta $|\bs{\kappa}|>2$. This behavior can be attributed to the short-range nature of the pseudopotential interaction. In contrast, for the Coulomb interaction, a deep minimum would be visible (see Ref.~\onlinecite{Yang-PhysRevLett.108.256807} for a comparison between short range interactions and the Coulomb interaction). 

\begin{figure*}
\includegraphics[width = 0.29\linewidth]{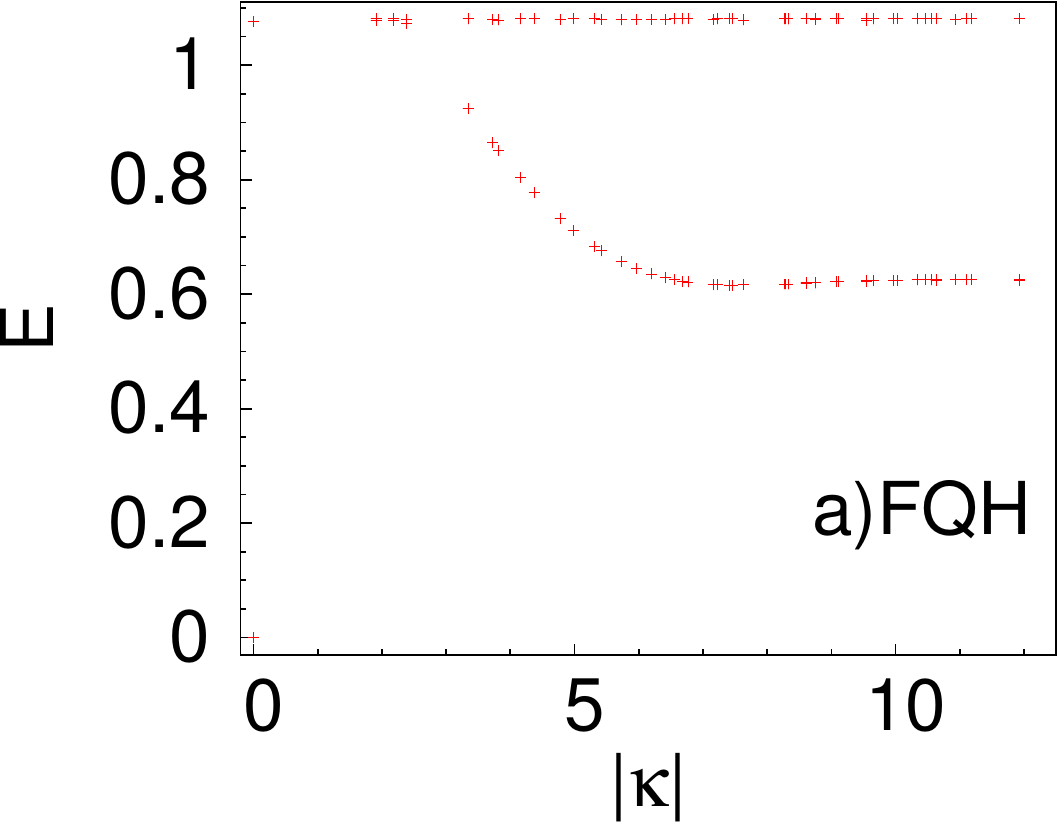}
\hspace{20pt}
\includegraphics[width = 0.29\linewidth]{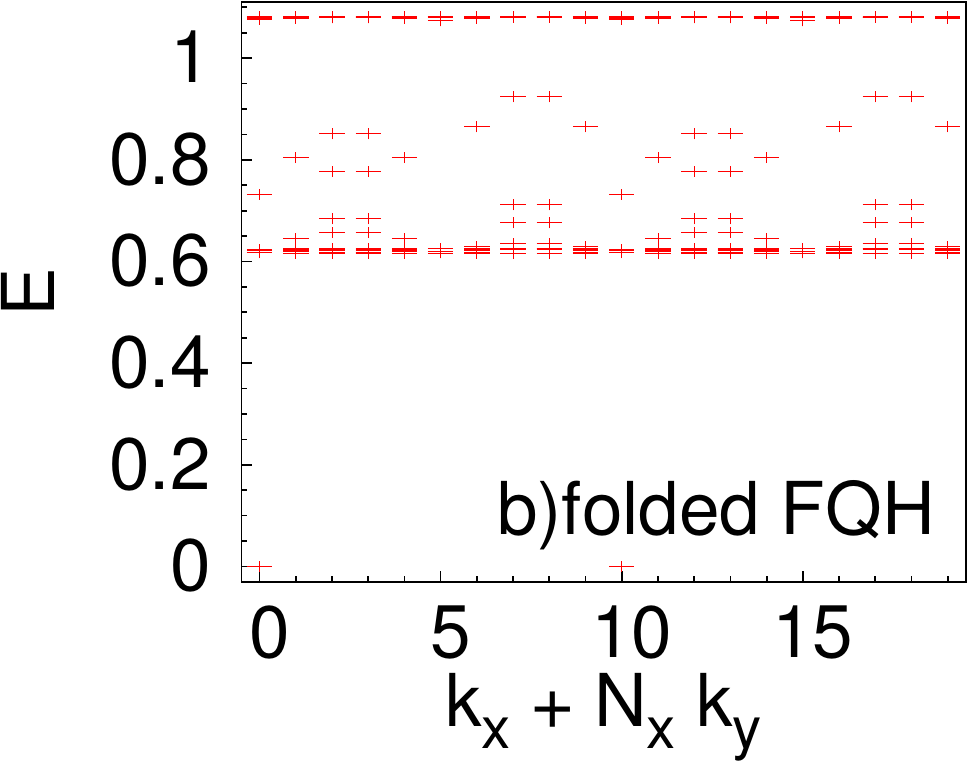}
\hspace{20pt}
\includegraphics[width = 0.3\linewidth]{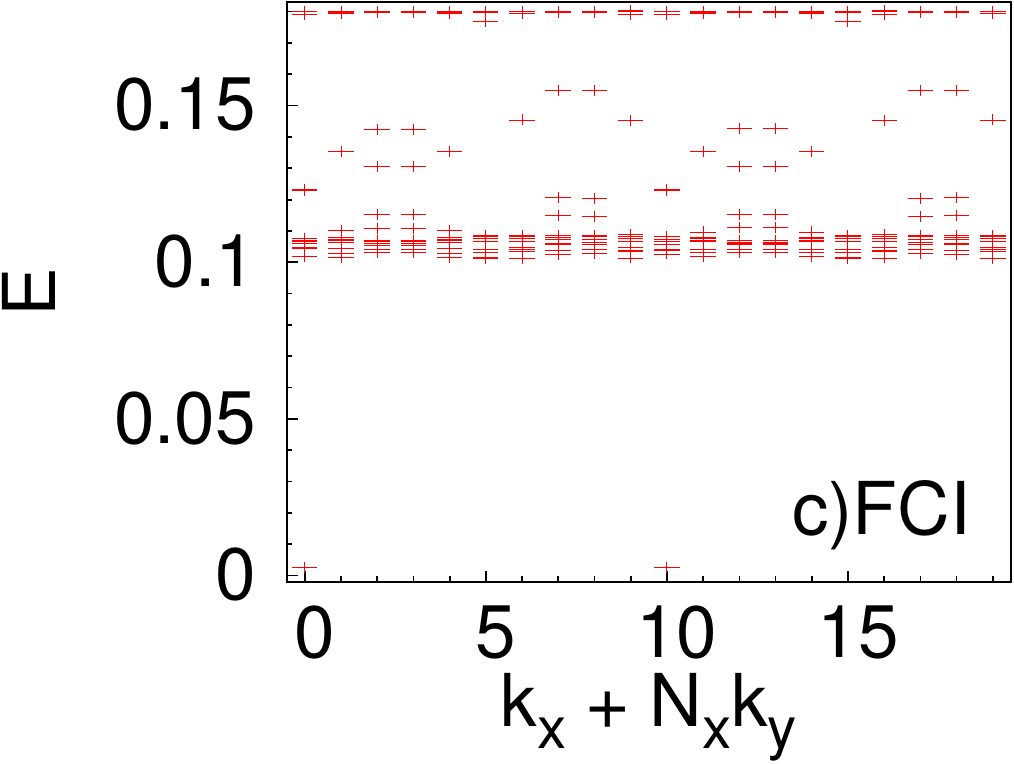}
\caption{Low energy spectrum of the FQH on a torus of aspect ratio $L_x/L_y = 1.25$ with a twisting angle $\theta = 2\pi/3$, with $N = 10$, $N_{\phi} = 20$, as a function of (a) the modulus of the momentum and (b) the linear FCI momentum (using the FQH-to-FCI mapping). (c) Low energy spectrum of the ruby lattice model with $N = 10$ particles and $N_x\times N_y = 5\times 4$.}
\label{fig:fqhTofciN10}
\end{figure*}

An elemental characterization of the magneto-roton mode is given by the number of states that it consists of in a finite size system. As the mode merges with the continuum for small momenta, there is an ambiguity in defining this number for certain system sizes. Here, we focus on systems where the number of magneto-roton states is unambiguous. 
We propose a simple phenomenological rule that determines how many states should be expected.
In the case of in the bosonic $\nu=1/2$ Laughlin state on the sphere geometry, there is no low energy excitation with an angular momentum $L = 0$ (where the ground state lies), nor in the sector with $L = 1$ (see Ref.~\onlinecite{Yang-PhysRevLett.108.256807} for an explanation in terms of the clustering properties of the Jack polynomials). By analogy, on the torus we expect no low energy excitation in the momentum sectors of the groundstate (which has a twofold degeneracy), nor in the sector with the smallest non-zero $|\bs{\mathsf{k}}|$ [for instance $\bs{\mathsf{k}} = (1,0)^{\mathsf{T}}$]. 
For a torus with a twisting angle $\theta = \pi/2$ and aspect ratio 1, the latter momentum sector has a degeneracy $4m$ due to the $C_4$ rotational symmetry and the CM translational symmetry. As a result, only $m(N^2 - 1 - 4)$ momentum sectors out of a total number of $NN_\phi$ momentum sectors in $\mathrm{BZ}_{\mathrm{FQH}}$ contribute a state to the magneto-roton mode. 
Note that the geometry of the torus has a crucial influence on the number of states. For instance, a torus with a twisting angle $\theta = 2\pi/3$ has $C_6$ symmetry. This, when combined with the CM translational symmetry, implies a $(6m)$-fold degeneracy of the sector $\bs{\mathsf{k}}=(1,0)^{\mathsf{T}}$. Hence, we expect to find magneto-roton states in $m(N^2 - 1 - 6)$ momentum sectors only.
We confirmed these counting rules in all cases that we analyzed. For instance, a system with $N = 10$ bosons at filling $\nu=1/2$ has a magneto-roton mode with $190$ states and $186$ states for $\theta = \pi/2$ and $\theta = 2\pi/3$, respectively. 
This counting of magneto-roton excitations should be contrasted to the counting of topological charged excitations (i.e., quasiholes and quasiparticles), which is independent of the geometry parametrized by $\theta$.

\subsection{Fractional quantum Hall to fractional Chern insulator folding}
\label{sec:fqhFolding}
A FCI emerges in a Chern insulator defined by a given tight-binding model if one partially fills a topologically non-trivial band with interacting fermions or bosons. For certain models and interactions, FQH-like phases emerge at specific filling factors. We consider a system with $N_x$ (respectively $N_y$) unit cells in the $x$ (respectively $y$) direction and periodic boundary conditions in both directions. 
For the FCI, translation operators in the $x$ and $y$ directions commute with each other and with the Hamiltonian. The eigenstates are labelled by $N_x\times N_y$ momentum quantum numbers $\bs{k}\in \mathrm{BZ}_{\mathrm{FCI}}$ with
\be
\begin{split}
\mathrm{BZ}_{\mathrm{FCI}}
\equiv&\left\{
\bs{k}=\frac{2\pi}{a_xN_x}\mathsf{k}_x \bs{\tilde{e}}_x + \frac{2\pi}{a_yN_y}\mathsf{k}_y \bs{\tilde{e}}_y
\right|
\\
&\left.
\mathsf{k}_x=0,\cdots, N_x-1;
\
\mathsf{k}_y=0,\cdots, N_y-1
\right\},
\end{split}
\label{eq: FCI BZ}
\ee
where $a_x$ and $a_y$ are the lattice spacings in the $x$- and $y$-direction, respectively.
The FCI has to be compared to a FQH system with $N_{\phi} = N_x\times N_y$, so that the number of single-particle states in a nondegenerate band on the lattice equals the number of orbitals in a Landau level. 
However due to the CM translational symmetry, the relative momenta of the FQH system reside in an $N\times N$ reduced Brillouine zone $\mathrm{BZ}_{\mathrm{FQH}}^{\mathrm{red}}$. A mapping between the $N^2$ FQH momenta and the $N_\phi$ lattice momenta, which corresponds to the folding of $\mathrm{BZ}_{\mathrm{FQH}}^{\mathrm{red}}$  down to $\mathrm{BZ}_{\mathrm{FCI}}$, was developed in Ref.~\onlinecite{Bernevig-2012PhysRevB.85.075128}.
Following this procedure, we show the folded FQH magneto-roton spectrum for $N = 10$ bosons in Fig.~\ref{fig:fqhTofciN10}(b). We now focus on the fate of the FQH magneto-roton mode under this mapping. In this representation, the magneto-roton mode consists of a highly degenerate low energy band, with a few states lying in the gap above the band. These more isolated states constitute the dispersive branch of the mode. The folding places states with a short and a long wavelength (in the FQH sense) in the same sectors. This obscures the identification of the magneto-roton mode as a single dispersing branch of states. 

This demonstrates the difficulties we will face to identify the dispersion relation of a potential magneto-roton mode in a FCI spectrum. Generically, FCIs do not have a CM translation symmetry that makes $\mathsf{k}_x$ a good quantum number in the FQH case. As a result, the FCI spectrum cannot be unfolded or resolved in this extra quantum number, and cannot be plotted as a function of $|\bs{k}|$ with $\bs{k}\in \mathrm{BZ}_{\mathrm{FQH}}^{\mathrm{red}}$.

\subsection{Fractional Chern insulators}
\label{sec:FCISMAConstruction}

\begin{figure}
\includegraphics[width = 0.48\linewidth]{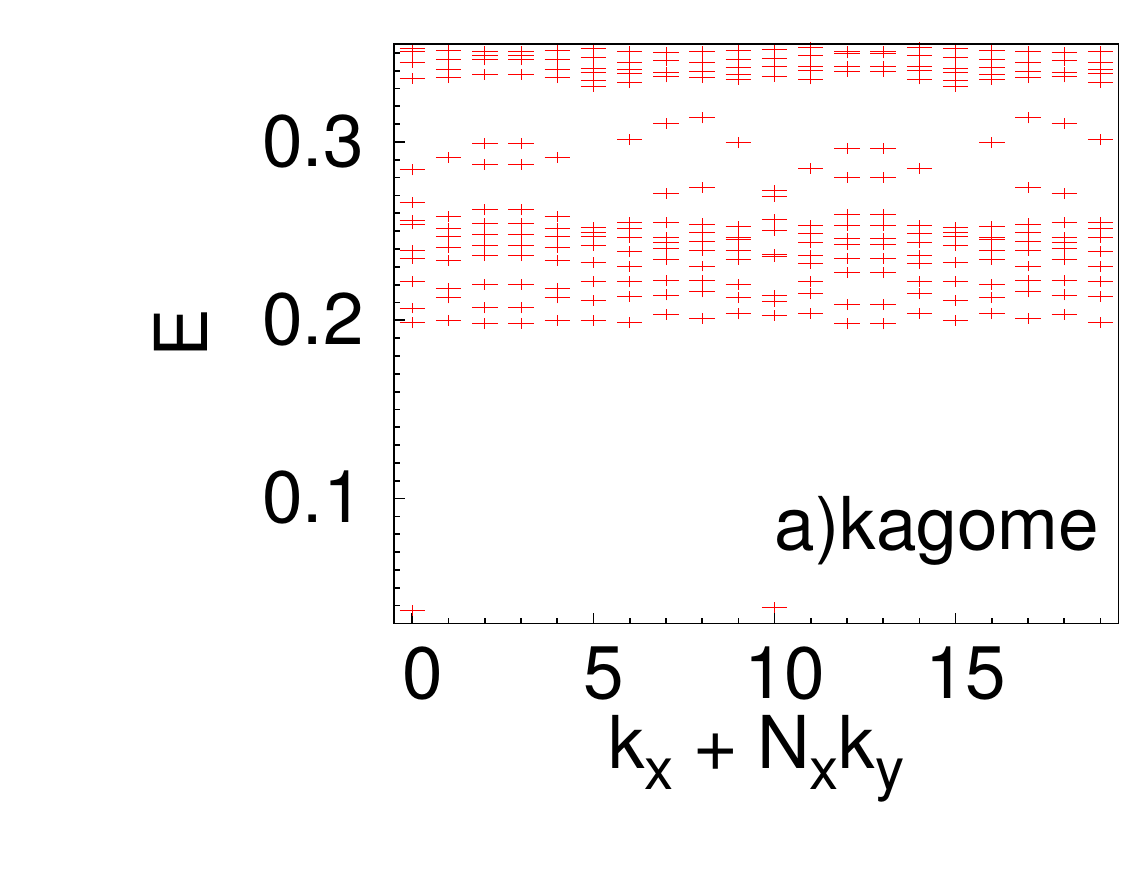}
\includegraphics[width = 0.48\linewidth]{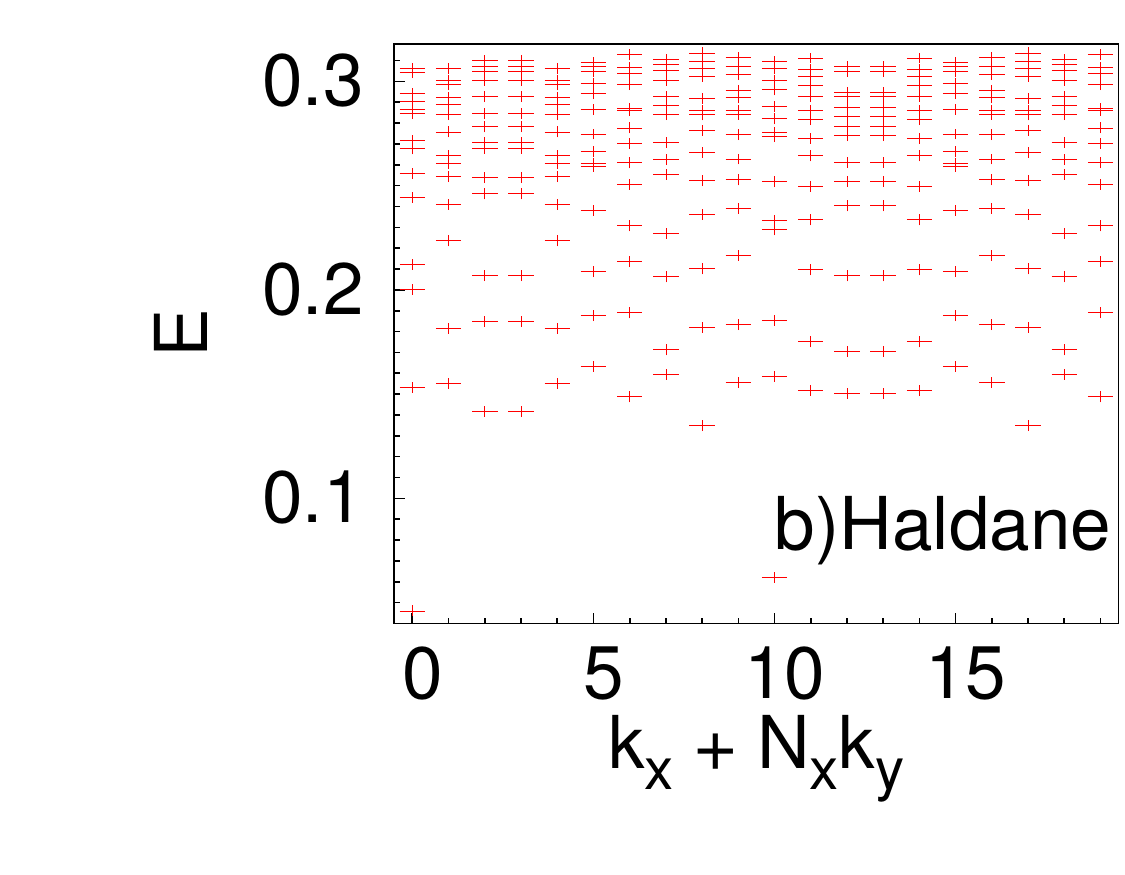}
\caption{Low energy spectrum of (a) the kagome lattice model and (b) the Haldane model with $N = 10$ bosons on a $N_x\times N_y = 5\times 4$ lattice. For the ruby lattice model [see Fig.\ref{fig:fqhTofciN10}(c)] and kagome lattice model, the ground state's quasi-degeneracy is such that no splitting is visible by the naked eye. The low energy excitation mode is separated by a gap from the continuum of higher energy excitations. In the case of the Haldane model, the ground state has a clear energy splitting, and the low energy mode cannot be distinguished from the continuum.}
\label{fig:FCISpectra8_9_10}
\end{figure}

Unlike in the case of the FQH effect with pseudopotential interaction, there exists no ``canonical'' model for FCIs without continuously tunable parameters. Rather, many details of FCI states are governed by nonuniversal aspects of the respective models.   
In this work, we consider three representative models for Chern insulators: the ruby~\cite{hu-PhysRevB.84.155116} and kagome~\cite{tang-PhysRevLett.106.236802} lattice models, and the Haldane model~\cite{haldane-1988PhRvL..61.2015H}. All three models have a lowest Bloch band characterized by a Chern number $1$. The tight binding parameters that we use are defined in 
Ref.~\onlinecite{Liu-2013PhysRevB.87.205136} (ruby lattice model),
Ref.~\onlinecite{repellin-2014arXiv1402.2652R} (kagome lattice model) and 
Ref.~\onlinecite{dobardzic-PhysRevB.88.115117} (Haldane model). We consider $N$ bosons on a $N_x\times N_y$ lattice with periodic boundary conditions. They interact via an on-site density-density interaction, which is projected onto the lowest band. The filling fraction is defined with respect to the lowest band, i.e., $\n = N/\left(N_x N_y \right)$, and chosen to be $\nu=1/2$ for all numerical calculations in this paper. It has been established that the ground state of these systems is a Laughlin-like phase: In the exact diagonalization spectra, we observe an almost degenerate twofold ground state with a gap to higher energy excitations. In the cases of the ruby and kagome lattice models, the ground state energy splitting is barely noticeable, proving that they are less affected by finite-size effects. We observe a low energy excitation mode separated in energy from the continuum of higher energy excitations [see Fig.~\ref{fig:fqhTofciN10}~(c)and Fig.~\ref{fig:FCISpectra8_9_10}~(a)]. This mode resembles the magneto-roton mode of 
the FQHE on the torus folded onto the FCI Brillouin zone [see Fig.~\ref{fig:fqhTofciN10}~(b)].
In the case of the Haldane model, there is a clear ground state splitting and no low energy excitation mode is distinguishable from the continuum [see Fig.~\ref{fig:FCISpectra8_9_10}~(b)]. The energy fluctuations that we see in the ground state are also present in the low energy excitations, resulting in their mixing with the continuum. This strong model dependency stresses the importance of choosing a ``good'' FCI model to observe the magneto-roton mode. Here, the qualifier ``good'' simultaneously applies to the optimization of ground state splitting, the gap in the entanglement spectrum, the energy gap, and a clear magneto-roton mode. All of these qualities seem to go hand in hand in the models we studied so far.

We compare the number of states in the magneto-roton mode in the FQH and FCI systems using the FQH-to-FCI mapping~\cite{Bernevig-2012PhysRevB.85.075128} (see Sec~\ref{sec:fqhFolding}). The counting per momentum sector in $\mathrm{BZ}_{\mathrm{FCI}}$ is the same provided the twisting angle of the torus $\a$ matches the angle defined by the reciprocal lattice vectors of the FCI ($\theta = \pi/3$ for the kagome lattice, $\theta = 2\pi/3$ for the ruby lattice). We compare the FCI spectrum to the FQH spectrum (folded into the FCI Brillouin zone), at the same system size and aspect ratio, identifying a similar magneto-roton pattern in the FCI as in the FQH spectra. An almost degenerate band including a large number of states lies below a few isolated states. In the FQH case, the isolated states were part of the dispersive branch of the magneto-roton mode at low momentum. In the case of the ruby lattice model, one can establish a one-to-one FQH-FCI correspondence of these states [see Fig.~\ref{fig:fqhTofciN10}(b) and (c)]. The near-degeneracy of some of these states 
comes from some residual FQH symmetry. In the kagome lattice model, this near-degeneracy is lifted into a low lying band that mixes with other states. Once again, we see that choosing a good model is crucial to observe the magneto-roton mode. The energy splitting widens the band of almost degenerate states to the point that they mix with the states of the dispersive branch, making their identification impossible.

\subsection{Extrapolation of the energy gap}
\label{sec:Gap Extrapolation}

\begin{figure*}[t]
\includegraphics[width = 0.32\linewidth]{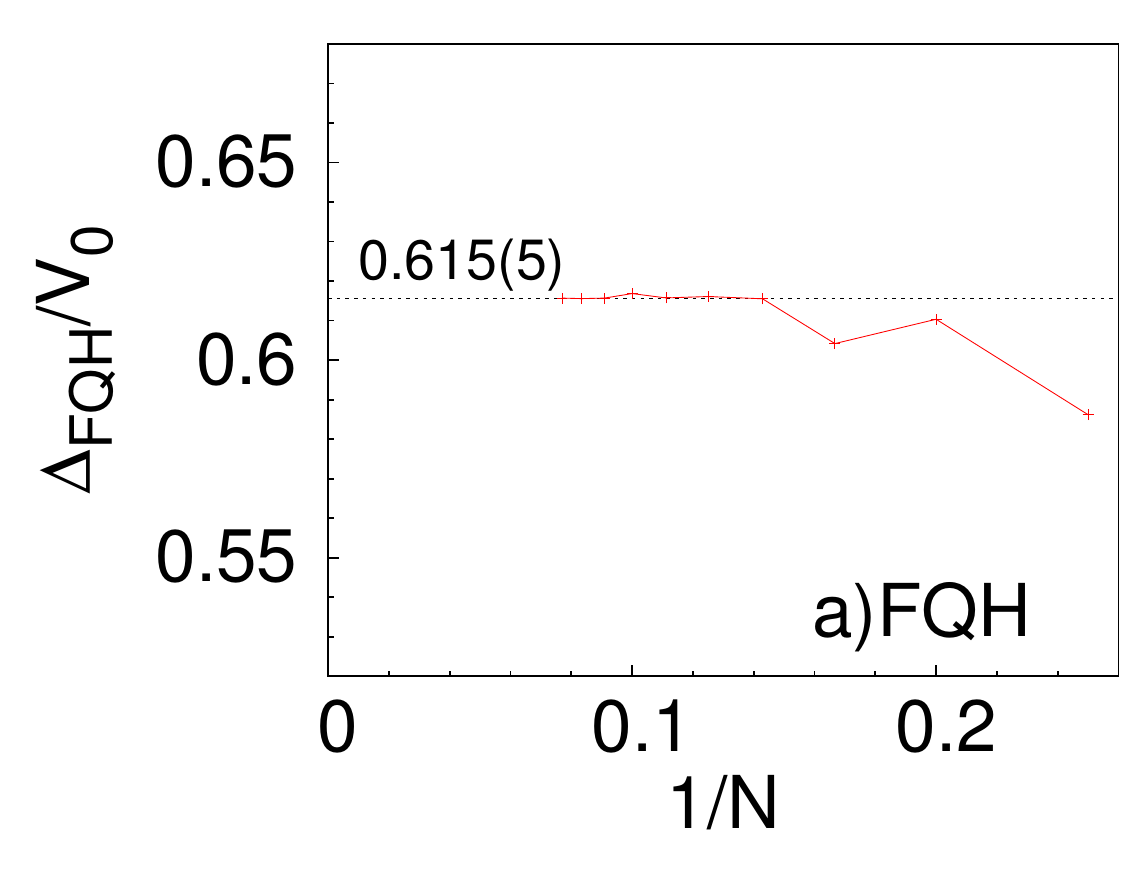}
\includegraphics[width = 0.32\linewidth]{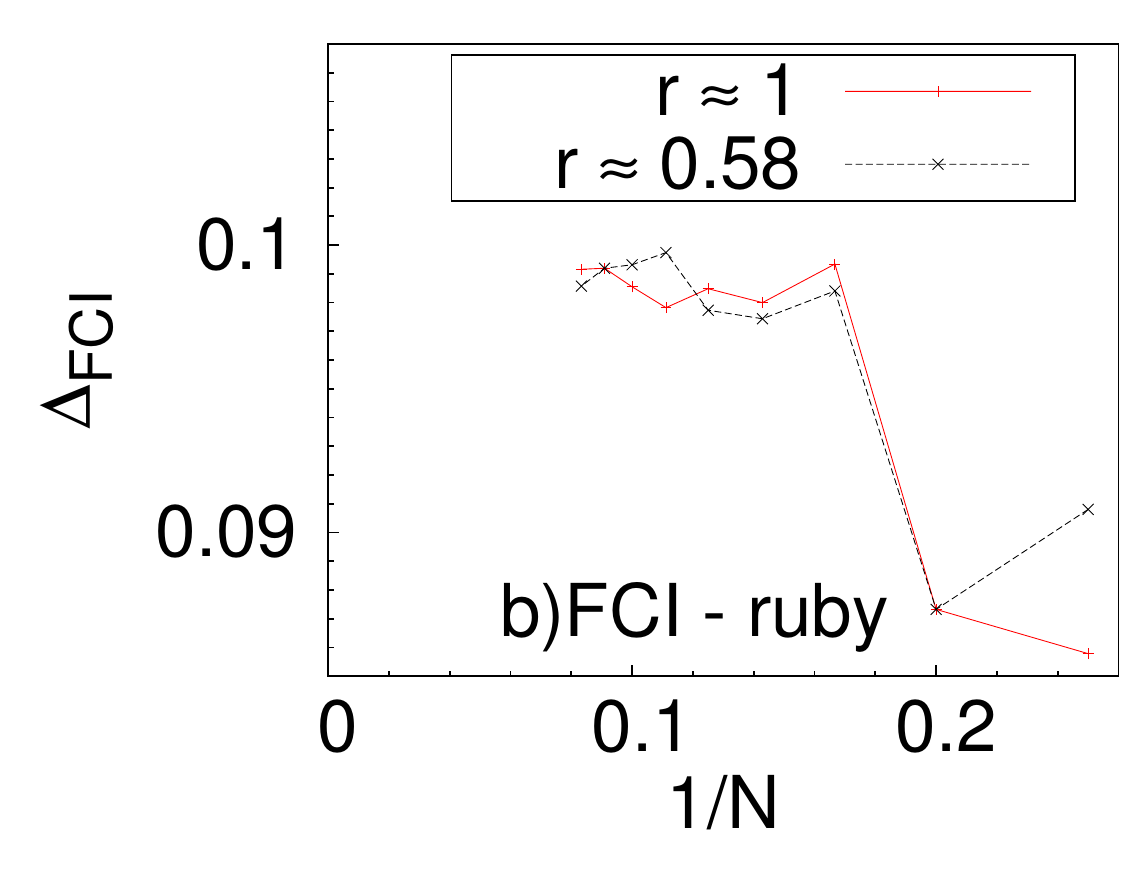}
\includegraphics[width = 0.32\linewidth]{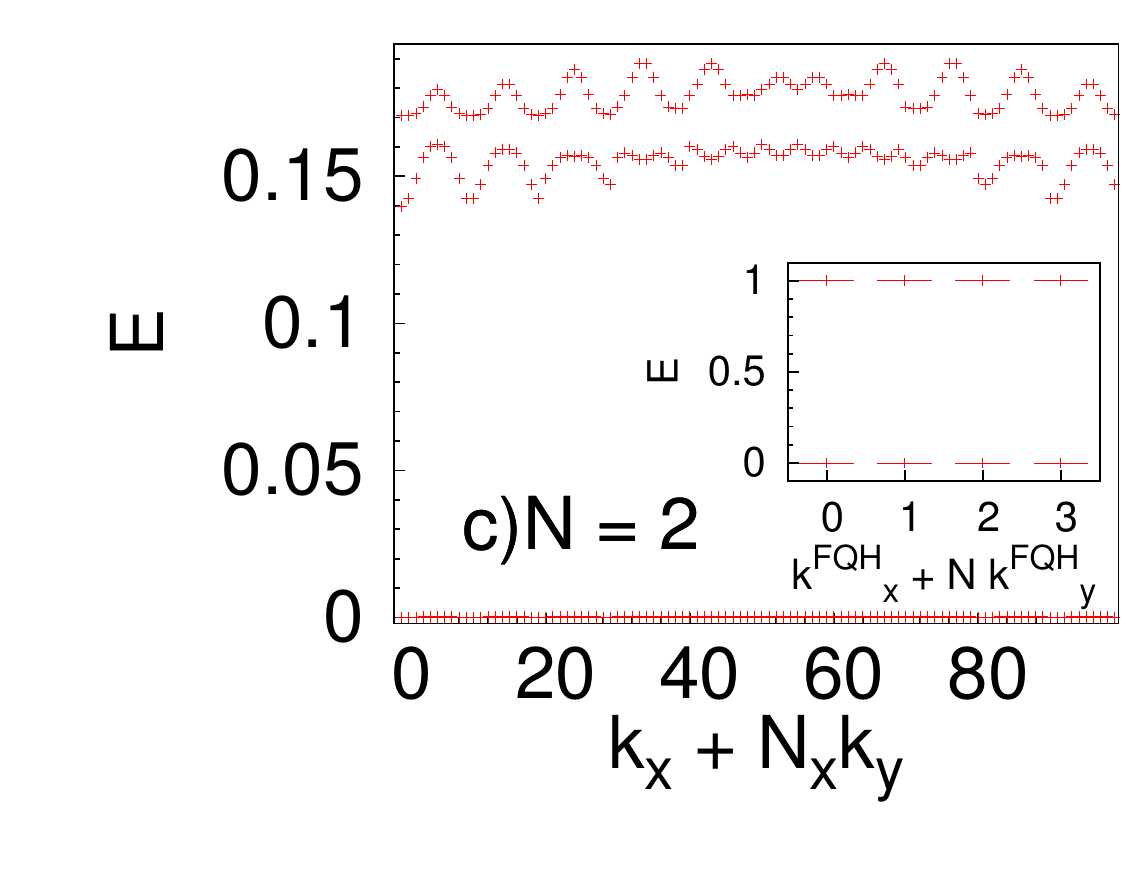}
\caption{
a) Gap of the FQH system on a torus of aspect ratio $1$, at filling factor $\n = 1/2$. 
b) Gap of the ruby lattice Chern insulator model at a filling factor $\n = 1/2$, for systems of aspect ratio as close to $1$ as possible, in one case, and as close to $0.58$ as possible.
c) Two-particle spectrum of the ruby lattice FCI model with on-site interaction, with $N_s = 10\times 10$ unit cells. The inset shows the two-particle spectrum of the FQH system on a torus with $N_{\phi} = 100$, and displayed in the reduced $2\times 2$ Brillouin zone.
}
\label{fig:GapScaling}
\end{figure*}

For a wide range of system sizes, the FQH magneto-roton modes all fall on the same curve, as shown Fig.~\ref{fig:magnetoroton}. Indeed, the gap between the ground state and the first excited state (which belongs to the magneto-roton mode) exhibits almost no finite size effect, starting from $N = 7$.
We extract the many body gap of the FQH systems and plot it as a function of $1/N$ in Fig.~\ref{fig:GapScaling} (a). The thermodynamic extrapolation of the gap yields a value of $\D = 0.615(5) \times V_0$, where the $V_0$ pseudo-potential is the scale of the two-particle interaction energy [consult the inset of Fig.~\ref{fig:GapScaling}~(c) for the definition of $V_0$]. Note that all the FQH energies are expressed in units of $V_0$ in this paper. The scaling of the gap on the sphere geometry was studied in Refs.~\onlinecite{regnault-PhysRevLett.91.030402,regnault-PhysRevB.69.235309}, and shows a more important finite size effect than we observe on the torus. The extrapolated gap of the Laughlin $\n = 1/2$ system on the sphere is $0.60(1) \times V_0$, in agreement with our value on the torus geometry. Note that the scaling on the sphere assumed a linear behavior as a function of $1/N$, which may underestimate the thermodynamic value of the gap.

Performing a similar extrapolation for a FCI system is more difficult. As was initially pointed out in Ref.~\onlinecite{regnault-PhysRevX.1.021014}, and discussed in great details in Refs.~\onlinecite{Bernevig-2012arXiv1204.5682B} and~\onlinecite{Lauchli-PhysRevLett.111.126802}, both extents $L_x$ and $L_y$ of the lattice should be large enough to prevent the formation of a charge density wave phase. As a consequence for a finite and generally small system size,  the aspect ratio of the lattice greatly influences the value of the gap. In order to minimize this effect while studying the evolution of the gap with the system size, we follow the approach introduced in Ref.~\onlinecite{Lauchli-PhysRevLett.111.126802}, and use tilted boundary conditions (see also Ref.~\onlinecite{repellin-2014arXiv1402.2652R} for a more detailed description). 
For any number of unit cells $N_s$, this method allows us to obtain an aspect ratio close to the desired value. We look at the evolution of the gap for systems with aspect ratios $r$ close to 1 and for systems $r \simeq 0.58$ in Fig.~\ref{fig:GapScaling}b. The choice of the value $r \simeq 0.58$ is rather arbitrary; it corresponds to the largest non-tilted lattice that we can numerically reach, namely with $N = 12$ particles ($N_s = 24 = 6 \times 4$). 
While one expects that $r \simeq 1$ should minimize the finite size effects, our system actually shows a small size dependence at either aspect ratio. 
Note that the energy scale of the interaction in the FCI is not as well defined as in the FQH case. The energy scale is parametrized by the gap in the spectrum of the interacting two-particle problem. While the FQH case only has one single non-zero energy per momentum sector, which is almost $\bs{k}$ independent, the FCI system has non-zero energy states with larger fluctuations, originating from the momentum-dependence of the projection on a given band [see Fig.~\ref{fig:GapScaling}~(c)]. This prevents us from performing an exact rescaling of the FCI spectra with respect to the two-particle energy scale. Nevertheless, taking the average two particle non-zero energy as the energy scale leads to an extrapolated gap of $0.60(3)$, a value close to the FQH one.

\section{Single mode approximation}
\label{ref: SMA}

Throughout this work, we consider the Laughlin state at filling fraction $\n = 1/m$. The SMA provides a variational expression for low energy excitations 
above the topologically degenerate ground states $\ket{\Psi_\alpha}$, $\alpha=0,\cdots,m - 1$,
given by
\be
\ket{\Psi_{\bs{k},\alpha}^{\mathrm{SMA}}} = \rho_{\bs{k}} \ket{\Psi_\alpha}.
\label{eq:defSMA}
\ee
Here, $\rho_{\bs k}$ is the Fourier component of the density operator at momentum $\bs{k}$ projected to the lowest Landau level (LLL) or any given Bloch band in the FCI case. The momentum of $\ket{\Psi_{\bs{k},\alpha}}$ is given by the momentum $\bs{K}_\alpha$ of $\ket{\Psi_\alpha}$ shifted by $\bs{k}$. As illustrated in the previous section, the FQH states possess a larger set of good momentum quantum numbers $\bs{k}$ than the FCI. In fact, a naive implementation of Eq.~\eqref{eq:defSMA} produces more variational SMA states than the number observed in the magneto-roton mode for the FQH effect. For the FCI, in contrast, the very definition of the density operator $\rho_{\bs{k}}$ itself is ambiguous. The FCI density operator depends parametrically on the geometrical embedding of orbitals in the unit cell of the underlying lattice. It will be the objective of this section to interpret Eq.~\eqref{eq:defSMA} correctly for both the FQH effect on the torus and for the FCI.

Given the states $\ket{\Psi_{\bs{k},\alpha}^{\mathrm{SMA}}}$, one can obtain an approximation to the dispersion of the  magneto-roton mode via
\be
E^{\mathrm{mr}}_{\bs{k}}
=\frac{\bra{\Psi_{\bs{k},\alpha}^{\mathrm{SMA}}}H\ket{\Psi_{\bs{k},\alpha}^{\mathrm{SMA}}}}{
\left\langle\Psi_{\bs{k},\alpha}^{\mathrm{SMA}}\right.\ket{\Psi_{\bs{k},\alpha}^{\mathrm{SMA}}}
},
\label{eq: variational SMA energy}
\ee
where $H$ is the many-body Hamiltonian of the FQH or FCI system.

\subsection{Fractional quantum Hall effect on the torus}
\label{sec:FQHSMATheory}

First, we introduce the expressions for the SMA in the case of the FQH. We consider a torus $\mathbb{T}=[0,L_x]\times[0,L_y]$ spanned by the two orthogonal unit vectors $\bs{e}_x$ and $\bs{e}_y$ that is pierced by $N_{\phi}$ flux quanta. 
The position space representation of a basis of single-particle wave states in the Landau gauge, and in the LLL, is given by
\ba
\phi_{\mathsf{q}_y}(\bs{r}) =  \frac{e^{-\frac{x^2}{2\ell_B^2}}}{\sqrt{\sqrt \pi L_y \ell_B}} \sum_{\mathsf{k}\in \mathbb Z} & & \biggl[ e^{\frac{2\pi}{L_y}(\mathsf{q}_y + \mathsf{k} N_{\phi})(x + iy)}
 \\
 & & \times e^{-\frac{1}{2}(\frac{2\pi \ell_B}{L_y})^2 (\mathsf{q}_y + \mathsf{k}N_{\phi})^2} \biggr]\, ,
\nonumber
\label{eq:LLLwavefunction}
\ea
where $\mathsf{q}_y =0,\cdots, N_{\phi} - 1$ is the conserved one-body momentum along the $y$ axis, $\bs{r}=(x,y)\in\mathbb{T}$, and the magnetic length is given by $\ell_B^2=\sin\theta L_xL_y/(2\pi N_\phi)$.
Using these wave functions, the density operator at position $\bs{r}$, when projected to the LLL, is expressed as 
\ba
\rho(\bs{r}) 
& = & \sum_{\mathsf{q}_{y}, \mathsf{q}_{y}^\prime} \phi^*_{\mathsf{q}_{y}}(\bs{r}) \phi_{\mathsf{q}_{y}^\prime}(\bs{r}) c^{\dagger}_{\mathsf{q}_{y}}c_{\mathsf{q}_{y}^\prime} .
\ea
Here, $c^{\dagger}_{\mathsf{q}_{y}}$ and $c_{\mathsf{q}_{y}}$ are the operators that, respectively, create and annihilate a particle in the orbital $\mathsf{q}_y$ of the LLL.
The Fourier components of the projected density operator are given by
\begin{subequations}
\be
\rho_{\bs{k}} = \int_{\mathbb{T}}\mathrm{d}^2\bs{r} \,e^{-i\bs{ k} \cdot \bs{r}} \rtilde(\bs{ r}),
\ee
and can be conveniently expressed as 
\begin{equation}
\rho_{\bs k} = e^{-\frac{\bs{ k}^2 \ell_B^2}{4}}e^{-\frac{2\pi i \mathsf{k}_x \mathsf{k}_y}{2N_{\phi}}}\sum_{\mathsf{q}_y = 0}^{N_{\phi} - 1} e^{-\frac{2\pi i \mathsf{k}_x\mathsf{q}_y}{N_{\phi}}} c^{\dagger}_{\mathsf{q}_y + \mathsf{k}_y}c_{\mathsf{q}_y}
\label{eq:FQHProjectedDensityOperator}
\end{equation}
\end{subequations}
In Eq.~\eqref{eq:FQHProjectedDensityOperator}, we do not restrict $\bs{k}$ to belong to $\mathrm{BZ}_{\mathrm{FQH}}$, as
was pointed out in Ref.~\cite{Chamon-PhysRevB.86.195125, Murthy-PhysRevB.86.195146}. 
Rather, for every $\bs{k}\in \mathrm{BZ}_{\mathrm{FQH}}$, there exist several linearly independent operators 
$\rho_{\bs{k}+\bs{G}}$ with the reciprocal lattice vectors
\be
\bs{G}=2\pi\,N\left( \mathsf{G}_x/L_x,m \mathsf{G}_y/L_y\right)^{\mathsf{T}},
\qquad
\bs{\mathsf{G}}\in\mathbb{Z}^2.
\label{eq: Gs}
\ee
More precisely, $\rho_{\bs{k}+\bs{G}}=\rho_{\bs{k}+\bs{G}'}$
if there exists a pair of integers $(\Delta\mathsf{G}_x,\Delta\mathsf{G}_y)$ so that $\bs{\mathsf{G}}-\bs{\mathsf{G}}'=(m \Delta\mathsf{G}_x,\Delta\mathsf{G}_y)$.
This gives rise to $m$ distinct density operators $\rho_{\bs{k}+\bs{G}}$ for every is $\bs{k}\in \mathrm{BZ}_{\mathrm{FQH}}$.
Hence, the $\rho_{\bs k}$ span the same $N_\phi^2$-dimensional space of operators as the boson bilinears $c^{\dagger}_{\mathsf{q}_y^\prime}c_{\mathsf{q}_y}$, with $\mathsf{q}_y, \mathsf{q}_y^\prime=0,\cdots, N_\phi-1$. 
Acting with the $\rho_{\bs k}$ operators on the $m$-fold degenerate ground states according to Eq.~\eqref{eq:defSMA} thus yields a basis of $mN_{\phi}^2$ linearly independent variational states $\left\{\ket{\Psi^{\mathrm{SMA}}_{\bs{k}+\bs{G},\alpha}}\right\}$, spanning what we call the bilinear subspace. That is, for every of the $N\times N_\phi$ good quantum numbers $\bs{k}\in\mathrm{BZ}_{\mathrm{FQH}}$, we can build $m^2$ variational states with the help of the density operator~\eqref{eq:FQHProjectedDensityOperator}. Here, one factor of $m$ is due to the $m$ degenerate ground states labeled by $\alpha$ that one can act on, and a second factor of $m$ comes from the distinct shifts by reciprocal lattice vectors $\bs{G}$.

\begin{figure}
\includegraphics[width = 0.8\linewidth]{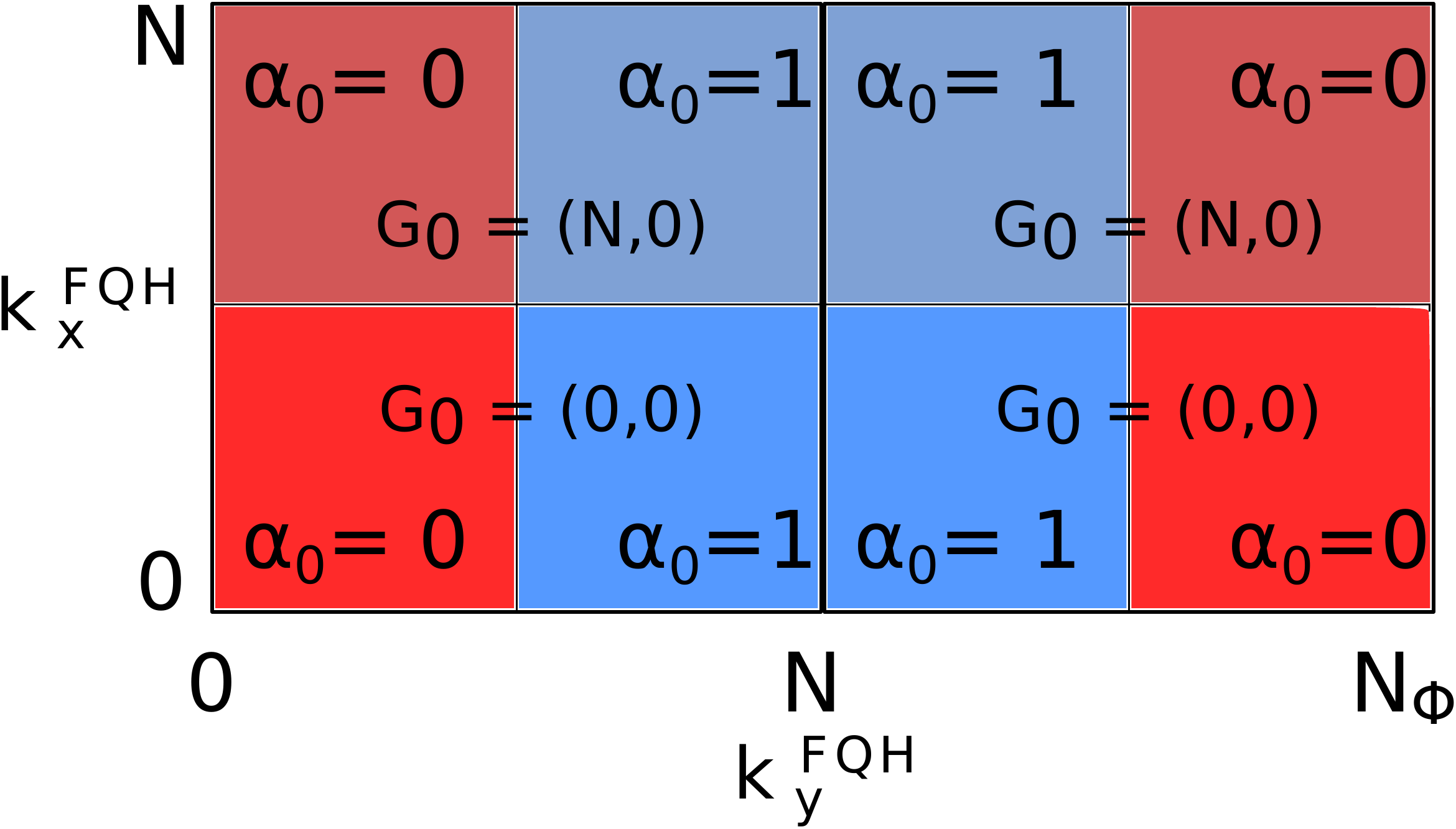}
\caption{Schematic representation of the SMA construction rule Eq.~\eqref{eq: distance in momentum space} in the FQH Brillouin zone $\mathrm{BZ}_{\mathrm{FQH}}$. In the vicinity of the origin (red), the magnetoroton states are well approximated by SMA states stemming from the first ground state ($\a_0 = 0$). Conversely, around $(\mathsf{k}_x, \mathsf{k}_y) = (0,N)$ (blue), the magneto-roton states are approximated by SMA states originating from the second ground state ($\a_0 = 1$). In the upper area, one should use the density operator at $\bs{k}+\bs{G}_0$ with $\bs{\mathsf{G}_0} = (N,0)$ to obtain the magneto-roton state. In the lower area, one should use the density operator at $\bs{k}+\bs{G}_0=\bs{k}$, where $\bs{k}\in \mathrm{BZ}_{\mathrm{FQH}}$.}
\label{fig:BZ}
\end{figure}

In contrast, we have seen in Sec.~\ref{ref:magnetoroton} that the magneto-roton mode consists of at most one state in every 
of the $N\times N_\phi$ sectors of $\bs{k}$.
Thus, the naive SMA as given by Eq.~\eqref{eq:defSMA} provides a factor of $m^2$ more variational states than needed to describe the magneto-roton mode. 
For each $\bs{k}\in\mathrm{BZ}_{\mathrm{FQH}}$, we propose the following rule to select one of the $m^2$ SMA states 
\begin{subequations}
\label{eq: criterion for FQH SMA}
\be
\ket{\Psi^{\mathrm{SMA}}_{\bs{k}+\bs{G},\alpha}}
\equiv\rho_{\bs{k}+\bs{G}}\ket{\Psi_\alpha},
\qquad \alpha=0,\cdots, m-1,
\label{eq: full SMA state subspace FQH}
\ee
 as the variational state for the magneto-roton mode:
 The variational magneto-roton state is given by
\be
\ket{\Psi_{\bs{k}}^{\mathrm{mr-SMA}}}
=
\ket{\Psi^{\mathrm{SMA}}_{\bs{k}+\bs{G}_0,\alpha_0}},
\ee 
where $\alpha_0$ labels the ground state at momentum $\bs{K}_\alpha\in \mathrm{BZ}_{\mathrm{FQH}}$
and $\bs{G}_0$ is the reciprocal lattice vector
for which the momentum-space distance
\be
 |\bs{k}+\bs{G}-\bs{K}_\alpha|
\label{eq: distance in momentum space}
\ee
is minimized for fixed $\bs{k}\in\mathrm{BZ}_{\mathrm{FQH}}$.
An illustration of this selection rule for the simplest case of $m=2$ is given in Fig.~\ref{fig:BZ}.
\end{subequations}

\subsection{Fractional Chern Insulators}
\label{sec:FCISMATheory}

For fractional Chern insulators, we consider the following generic form of a translationally invariant one-body Hamiltonian on a lattice $\Lambda$ of $N_{\mathrm{s}}=N_x\times N_y$ sites with periodic boundary conditions
\be
H_{\mathrm{CI}} = \sum_{\bs{r},\bs{r}'\in\Lambda}\sum_{a, a'} c^{\dagger}_{\bs{r} a}h_{a a'}(\bs{r}-\bs{r}')c_{\bs{r}' a'} ,
\label{eq:genericHamiltonian}
\ee
where $c_{\bs{r} a}$ and $c^{\dagger}_{\bs{r}a}$ are the operators that annihilate and create, respectively, a particle on the orbital $a=0,\cdots,N_{\mathrm{b}}$ of lattice site $\bs{r}\in\Lambda$. We use the Fourier transform convention 
\begin{subequations}
\be
c_{\bs{ k}a} = \frac{1}{\sqrt{N_\mathrm{s}}}\sum_{\bs{r}\in\Lambda} e^{i\bs{k}\bs{r}}c_{\bs{r}a}.
\ee
Using these operators, the Hamiltonian~\eqref{eq:genericHamiltonian} is represented in terms of the Bloch Hamiltonian $h_{a a'}(\bs{k})$ as
\be
H_{\mathrm{CI}} = \sum_{\bs{ k}}\sum_{a a'}c^{\dagger}_{\bs{ k} a} h_{a a'}(\bs{ k}) c_{\bs{ k} a'}.
\ee
Here, $\bs{k}$ takes values in the FCI Brillouin zone $\mathrm{BZ}_{\mathrm{FCI}}$ that was defined in Eq.~\eqref{eq: FCI BZ}.
\end{subequations}

For every $\bs{k}\in \mathrm{BZ}_{\mathrm{FCI}}$, the Bloch Hamiltonian has a spectral decomposition into normal modes $\gamma^n_{\bs{k}}$ with band index $n=0,\cdots, N_{\mathrm{b}}$,
\begin{subequations}
\be
H_{\mathrm{CI}} = \sum_{\bs k}\sum_{n}E_{\bs{k},n} \gamma^{n\dagger}_{\bs k}\gamma^n_{\bs k}.
\ee
The normal modes are related by a unitary transformation to the operators $c_{\bs{ k}a}$
\be
\gamma^n_{\bs k} = \sum_{a'}u^{n *}_{\bs{k} a'}c_{\bs{ k} a'},
\ee
\end{subequations}
where the matrix elements $u^{n *}_{\mbf k a'}$ form the eigenstates of the Bloch Hamiltonian $h_{a a'}(\bs{ k})$.
To define a density operator (and subsequently project it to a given band $n=0$), 
a geometrical choice about the embedding of the orbitals $a=1,\cdots, N_{\mathrm{b}}$ has to be made by assigning 
a displacement vector $\bs{r}_a$ to every orbital that locates it relative to a fixed point in the unit cell.
 We define an embedding as the set  $\{ \bs{ r}_{a} \}_{N_{\mathrm{b}}}$ of these displacements. 
For instance, we choose the embedding of the kagome lattice model as
\ba
\{ \bs{ r}_{a} \}_{N_{\mathrm{b}}}=\{ (0,0), (1/2, 0), (0, 1/2) \},
\label{eq: Kagome embedding}
\ea
in units where the lattice spacing is unity. In the same units, the ruby lattice model has the embedding
\ba
\{ \bs{ r}_{a} \}_{N_{\mathrm{b}}}=\frac{1}{3+\sqrt{3}}
\left\{
\begin{array}{c c c c c}
 \Big(&-1 & , & \frac{-1 + \sqrt 3}{2}&\Big), \\
  \Big(&1 &, & \frac{1 + \sqrt 3}{2}&\Big), \\
  \Big(&-1-\sqrt 3& , & \frac{-1 -\sqrt 3}{2}&\Big), \\
  \Big(&1 & , & \frac{1-\sqrt 3}{2}&\Big), \\ 
 \Big(&-1 & , & \frac{-1 -\sqrt 3}{2}&\Big), \\
  \Big(&1 + \sqrt 3 &,& \frac{1+\sqrt 3}{2}&\Big) \\
\end{array}
\right\}.
\label{eq: ruby embedding}
\ea

\begin{figure}
\includegraphics[width = 0.8\linewidth]{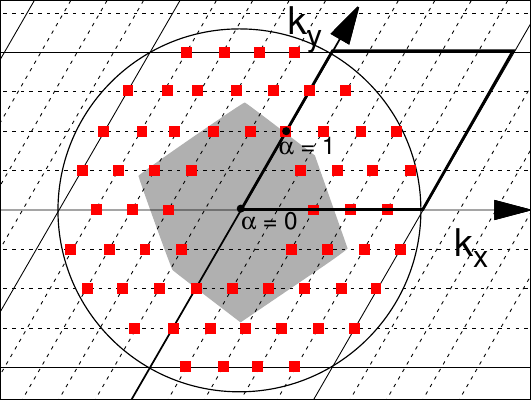}
\caption{Schematic representation of the SMA construction rule Eq.~\eqref{eq:  conduction with Kmax} for the ruby FCI model with $N = 10$ particles around the first ground state $\a = 0$. The dashed lines represent the reciprocal lattice, while the solid lines represent the limits of each Brillouin zone ($\bs{G}$ takes a different value in each of these zones). The bold line marks the limits of the first Brillouin zone $\mathrm{BZ}_{\mathrm{FCI}}$ ($\bs{G} = 0$), while the grey area corresponds to the first Brillouin zone when constructed as the Wigner-Seitz cell of the reciprocal lattice. The center of mass momentum of each of the two ground states is indicated by a black dot. We draw a red square for each SMA state satisfying $|\bs{k}+\bs{G}-\bs{K}_0|<K_{\mathrm{max}}$, with the position of $K_{\mathrm{max}}$ represented by a circle. Note that no SMA state is realized at $\bs{k} + \bs{G} = 0$ or the six sectors in its vicinity, similarly to the FQH case, and in agreement with the counting rule of Sec.~\ref{ref:magnetorotonFQH}.}
\label{fig:FCISMARule}
\end{figure}

The choice of embedding is an extra piece of information that is not contained in the Hamiltonian and determines whether or not the density operator shares certain spatial symmetries with the Hamiltonian. For example, the choice Eq.~\eqref{eq: Kagome embedding} preserves the inversion symmetry of the kagome lattice Hamiltonian. 
The density operator $\tilde{\rho}_{\bs{ k}}$, and its corresponding projection $\rho_{\bs k}$ in the band $n=0$ are given by
\begin{subequations}
\ba
\tilde{\rho}_{\bs k} & = & \sum_{\bs{r}\in \Lambda}\sum_{a} 
e^{i\bs{ k}\cdot (\bs{r} + \bs{r}_{a})} c^{\dagger}_{\bs{r}a}c_{\bs{r}a},
\\
\rho_{\bs k} & = & 
\sum_{\bs q\in\mathrm{BZ}_{\mathrm{FCI}}}
\left[
\sum_{a} 
e^{i\bs{k}\cdot\bs{ r}_{a}} u^{0 *}_{\bs{q} + \bs{k},a} u^{0}_{\bs{ q}, a} 
\right]
\gamma^{0\dagger}_{\bs{k}+\bs{q}}\gamma^{0}_{\bs q}.
\label{eq:embedding}
\ea
When $\bs{q} = \frac{2\pi}{N_{a}}\tilde{\bs{e}}_{x,y}$, the bracketed factor in Eq.\pref{eq:embedding} can be identified to the nonunitary exponentiated Abelian Berry connection ${\cal A}_{a}(\bs{k})$. Hence, ${\cal A}_{a}(\bs{k})$ also depends on the embedding, a piece of information not contained in the effective Hamiltonian. Still, as was discussed in Ref.~\cite{Wu-PhysRevLett.110.106802}, the embedding has to be properly chosen to obtain a large overlap of the model wavefunctions with exact diagonalization states. For any specific model, we will use the same embedding that maximizes the overlap with the model state. More precisely, we will use the embedding defined in Eq.~\eqref{eq: ruby embedding} for the ruby lattice model, and the embedding of Eq.~\eqref{eq: Kagome embedding} for the kagome lattice model.

For later use, we shall also define a variant of the projected density operator that involves the \emph{unitary} Berry connection
\be
\rho_{\bs k}^{\mathrm{U}(1)} = 
\frac{1}{N_{\mathrm{b}}}
\sum_{\bs q\in\mathrm{BZ}_{\mathrm{FCI}}}
\frac{\sum_{a} 
e^{i\bs{k}\cdot\bs{ r}_{a}} u^{0 *}_{\bs{q} + \bs{k},a}u^{0}_{\bs{ q}, a}}
{|\sum_{a} e^{i\bs{k}\cdot\bs{ r}_{a}}e^{i\bs{k}\cdot\bs{ r}_{a}} u^{0 *}_{\bs{q} + \bs{k},a}u^{0}_{\bs{ q}, a}|}
\gamma^{0\dagger}_{\bs{k}+\bs{q}}\gamma^{0}_{\bs q}.
\label{eq: U(1) density operator}
\ee
\end{subequations}
This definition of a density operator has proven useful to establish a mapping between FCI and FQH states~\cite{Wu-PhysRevLett.110.106802} and we shall see that it also produces slightly better results  for the SMA to the magneto-roton mode than $\rho_{\bs k}$.
Note that both $\rho_{\bs k}$ and $\rho_{\bs k}^{\mathrm{U}(1)}$ do not in general go back to themselves when $\bs{k}$ is shifted by a reciprocal lattice vector 
\be
\bs{G}=2\pi\left(\mathsf{G}_x/a_x,\mathsf{G}_y/a_y\right) ,
\qquad
\bs{\mathsf{G}}\in\mathbb{Z}^2,
\ee
if the embedding displacements $\bs{r}_a$ are not integer in units of the lattice spacing. Thus, $\bs k$ in Eq.~\eqref{eq:embedding} is not limited to $\mathrm{BZ}_{\mathrm{FCI}}$. In the case of the kagome lattice with the embedding~\eqref{eq: Kagome embedding}, 
$\rho_{\bs{k}+\bs{G}}=\rho_{\bs{k}+\bs{G}'}$ if $\bs{\mathsf{G}}-\bs{\mathsf{G}}'\,\mathrm{mod}\,2=0$, yielding $4N_{\mathrm{s}}$ independent density operators. 
The number of linearly independent density operators (i.e. the number of values of $\bs k$ that give linearly independent density operators) depends on the model's particular embedding. Consequently, there is an arbitrariness in using a specific embedding to obtain $\rho_{\bs k}$. 
For any incommensurate embedding, Eq.~\eqref{eq:embedding} will yield $N_{\mathrm{s}}^2$ linearly independent density operators, spanning the same space as the full set of bilinear operators $\{\g^\dagger_{\bs{q} + \bs{k}}\g_{\bs{q}}|{\bs{k}, \bs{q} \in \mathrm{BZ}_{\mathrm{FCI}}}\}$.
As with the case of the FQH effect, when these operators are applied to the $m$ topological ground states, 
Eq.~\eqref{eq:defSMA} yields a factor of $m^2$ more variational SMA states than the number of states we observe in the magneto-roton mode. However, in contrast to the FQH effect, more than one magneto-roton state can reside in a sector of given $\bs{k}\in \mathrm{BZ}_{\mathrm{FCI}}$.
We propose the following rule to build a set of good variational states for each $\bs{k}\in \mathrm{BZ}_{\mathrm{FCI}}$:
The variational magneto-roton states are given by
\begin{subequations}
\label{eq: criterion for FCI SMA}
\be
\ket{\Psi_{\bs{k},i}^{\mathrm{mr-SMA}}}
=
\ket{\Psi^{\mathrm{SMA}}_{\bs{k}+\bs{G}_i,\alpha_i}},
\ee 
where the index $i$ enumerates all ground states $\alpha_i$ at momentum $\bs{K}_\alpha\in \mathrm{BZ}_{\mathrm{FQH}}$
for all reciprocal lattice vectors $\bs{G}_i$ 
that satisfy
\be
 |\bs{k}+\bs{G}-\bs{K}_\alpha|<K_{\mathrm{max}},
\label{eq: conduction with Kmax}
\ee
Here, $K_{\mathrm{max}}$ is a cutoff that is not fixed a priori, but sets a scale that does not depend on the system size. We give a schematic representation of this constraint for the ruby system with $N = 10$ bosons in Fig.~\ref{fig:FCISMARule}. The number of pairs $(\alpha_i,\bs{G}_i)$ that satisfy Eq.~\eqref{eq: conduction with Kmax} depends on $\bs{k}$. The total number of states that obey Eq.~\eqref{eq: conduction with Kmax} scales linearly with $N$, even though the number of states in the magneto-roton mode scales like $N\times N_s$. The FQH SMA itself provides $N\times N_{\phi}$ variational states. In spite of this, it will become clear in the next section that the number of states \emph{accurately} described by the SMA is the same in the FQH and the FCI systems.
\end{subequations}

Owing to the lower translational symmetry and the higher degree of model dependence, the SMA to the magneto-roton mode for the FCI contains more free parameters, such as the embedding and the cutoff $K_\mathrm{max}$. However, as we shall see in the next section, it does not stand behind the SMA for the FQH magneto-roton mode, even quantitatively.

\section{Numerical test of the single-mode approximation}
\label{ref: SMA data}

For each of the two classes of systems, the FQH states and the FCIs, we now test the SMA numerically. 
To that end, we consider the following three benchmarks
\begin{enumerate}
\item[(i)]
How good is the agreement between the variational energy~\eqref{eq: variational SMA energy} of the SMA states selected via the criteria~\eqref{eq: criterion for FQH SMA} and~\eqref{eq: criterion for FCI SMA} with the exact dispersion of the magneto-roton states for the FQH and FCI cases, respectively?
\item[(ii)]
How large is the overlap of these selected SMA states with the exact magneto-roton states?
\item[(iii)]
Do (i) and (ii) improve significantly if the full space 
SMA states~\eqref{eq:defSMA} is considered, instead of the subset selected by the criteria~\eqref{eq: criterion for FQH SMA} and~\eqref{eq: criterion for FCI SMA}? This serves as a direct test of the criteria~\eqref{eq: criterion for FQH SMA} and~\eqref{eq: criterion for FCI SMA}.
\end{enumerate}

As announced, we will focus on bosonic systems at half filling $\nu=1/2$, where the ground state is the twofold degenerate Laughlin state (i.e. $m = 2$). For the FQH states, we will use a delta-function interaction, while for the FCI, we consider the ruby lattice model with the interaction and model parameters given in Ref.~\onlinecite{Liu-2013PhysRevB.87.205136}.

\subsection{Fractional quantum Hall effect on the torus}
\label{sec:FQHSMANumerics}

\begin{figure}[t]
\includegraphics[width = 0.85\linewidth]{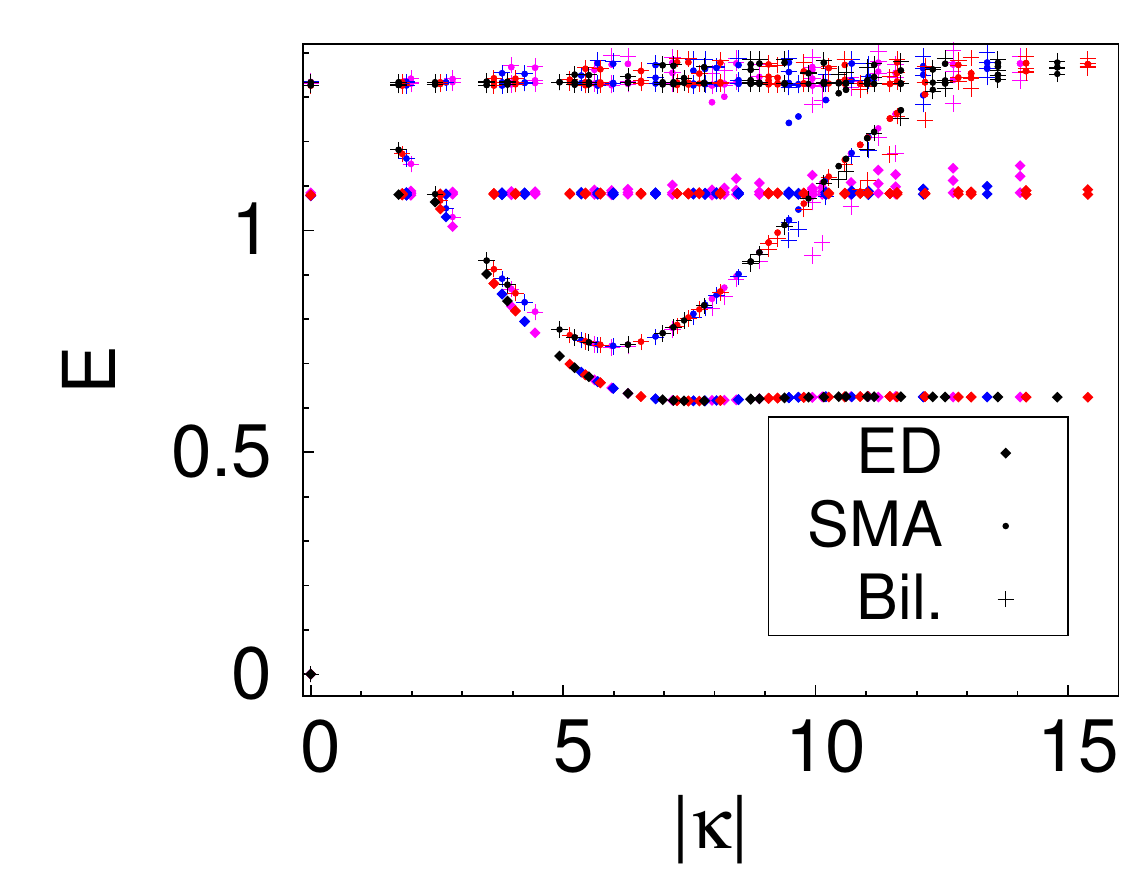}
\caption{Low energy spectrum of the FQH systems with up to $N = 13$ bosons on the torus, at half filling. The variational energy of the SMA states are compared to the energies obtained through exact diagonalization (ED) of the Hamiltonian, and to the energies obtained by diagonalization of the Hamiltonian in the full bilinear subspace (Bil.). For $N = 13$, only one eigenvalue per sector has been computed via ED. The color code for the systems sizes is the same as in Fig.~\ref{fig:FQHoverlap}.}
\label{fig:FQHEnergySMA}
\end{figure}

The numerical result that addresses the benchmark question (i) is summarized in Fig.~\ref{fig:FQHEnergySMA}, 
where data for various system sizes has been collapsed according to Eq.~\eqref{eq: K-rescaling for data collapse}. We observe that the SMA variational energies only slightly overestimate the magneto-roton mode energy at small momenta $|\bs{\kappa}|<2\pi$, in a way that accurately preserves the shape of the dispersion relation. 
For momenta $|\bs{\kappa}|>2\pi$, the SMA energies increase and finally merge with the continuum of excited states, while the magneto-roton mode flattens out to constant values. Note that neither the magneto-roton mode nor its approximation show any visible finite size effect. 
The estimated value of the excitation gap from the SMA dispersion is $0.74(2)$, which corresponds to a relative error of $0.20$ as compared to the exact diagonalization result.

\begin{figure}[t]
\includegraphics[width = 0.85\linewidth]{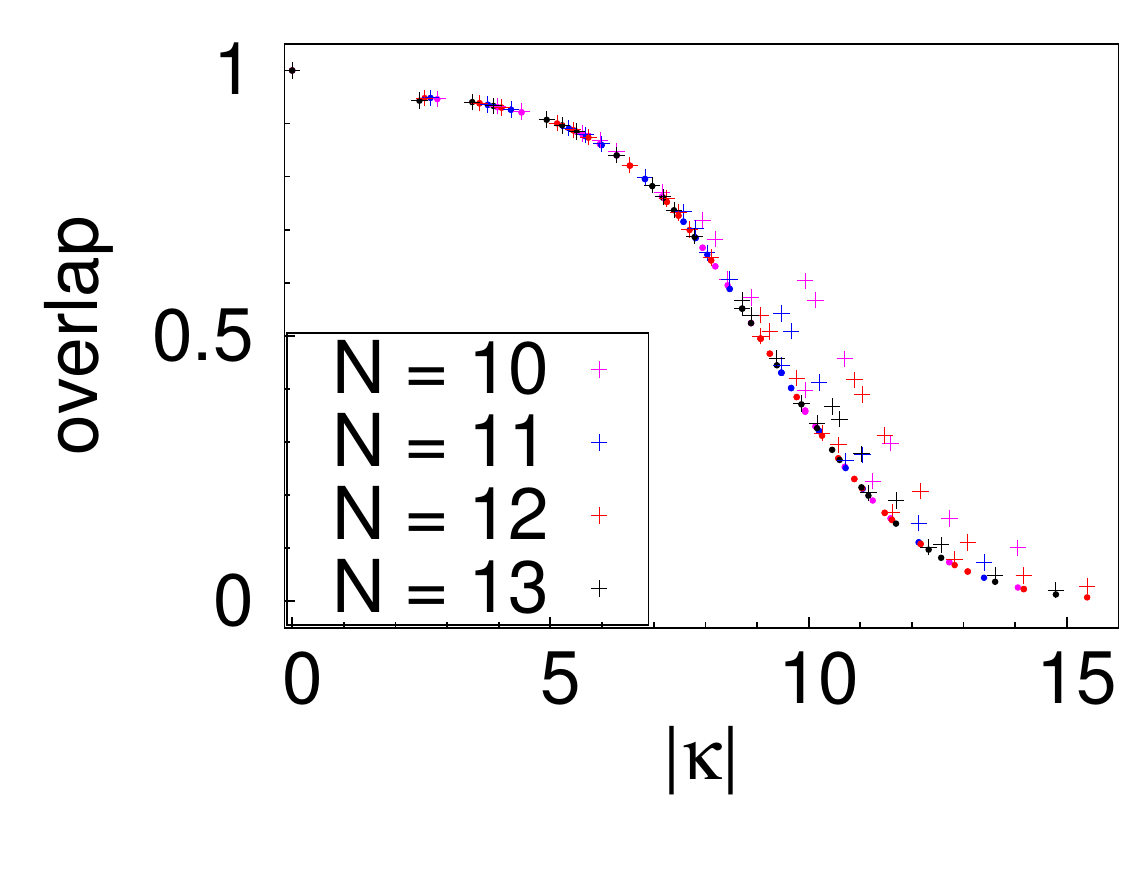}
\caption{The dots (respectively crosses) show the overlap between the SMA (respectively the full bilinear subspace) and each exact magneto-roton state, as a function of $|\bs{\kappa}|$, for the FQH on the torus with up to $N = 13$ bosons at half filling.}
\label{fig:FQHoverlap}
\end{figure}

Figure~\ref{fig:FQHoverlap} addresses benchmark question (ii) for the FQH effect. In accordance with the behaviour of the SMA variational energies, the overlap between the SMA state selected by criterion~\eqref{eq: criterion for FQH SMA} and the respective exact magneto-roton eigenstate at a given $\bs{k}\in\mathrm{BZ}_{\mathrm{FQH}}$ is high ($\simeq 0.9$) for small magnitudes of $|\bs{\kappa}|<2\pi$ of the rescaled momentum $\bs{\kappa}$, and drops significantly for $|\bs{\kappa}|>4\pi$. 

Finally, to address benchmark question (iii), and check the validity of the selection criterion~\eqref{eq: criterion for FQH SMA} for SMA states, we diagonalized the interacting Hamiltonian at every $\bs{k}\in\mathrm{BZ}_{\mathrm{FQH}}$ in the full $m^2$ subspace of SMA states~\eqref{eq: full SMA state subspace FQH}. 
The $m^2$ energy eigenvalues per momentum sector are superimposed with the exact and SMA spectra in Fig.~\ref{fig:FQHEnergySMA}. We observe that the enlarged space of variational states does not further improve the approximation to the magneto-roton dispersion that was obtained with the variational states selected by criterion~\eqref{eq: criterion for FQH SMA}. In particular, the flattening of the magneto-roton dispersion at large $|\bs{\kappa}|$ is not captured in this approach either. 
This is supported by the overlaps of the full SMA subspace with the magneto-roton mode being not significantly larger than the overlap of the single SMA state selected by criterion~\eqref{eq: criterion for FQH SMA} in each momentum sector (see Fig.~\ref{fig:FQHoverlap}). In the dispersive branch, the relative discrepancy between these overlaps is less than $10^{-4}$, while it is of the order of $0.2$ in the flatter part of the magneto-roton mode.
As the variational states~\eqref{eq: full SMA state subspace FQH} span the whole space of neutral single-particle excitations above the ground states, we conclude that the magneto-roton states are many-body excitations above the ground states at large  $|\bs{\kappa}|$.

In conclusion, we confirmed the validity of our selection criterion~\eqref{eq: criterion for FQH SMA} for variational magneto-roton states in the FQH effect on the torus and found that the SMA provides an excellent approximation to the magneto-roton mode for momenta $|\bs{\kappa}|<2\pi$, while it breaks down for $|\bs{\kappa}|>4\pi$. As a corollary, the number of magneto-roton states that are well captured by the SMA scales linearly with the number $N$ of particles in the system [see the definition of $\bs{\kappa}$ in Eq.~\eqref{eq: K-rescaling for data collapse}]. Remarkably, there is almost no finite size effect, and the magneto-roton mode is approximated with the same accuracy for any system size.

\subsection{Fractional Chern Insulators}
\label{sec:FCISMANumerics}

In evaluating the accuracy of the SMA to the magneto-roton mode of the FQH states, we were able to take advantage of the fact that states for which the SMA is a good or a poor approximation are naturally separated into small and large momenta 
$\bs{\kappa}$ in the Brillouin zone, respectively.
We illustrated in Fig.~\ref{fig:fqhTofciN10} that the spectrum of the FCI magneto-roton mode on the ruby lattice model can be very well 
reproduced by folding the spectrum of the FQH magneto-roton mode on the torus down to the Brillouin zone of the FCI. 
However, any attempt to mirror this approach for the SMA is obscured by the fact that under this folding the separation of small and large momenta is lost, because both small and large $\bs{k}\in\mathrm{BZ}_{\mathrm{FQH}}^{\mathrm{red}}$ may fall on the same $\bs{k}\in\mathrm{BZ}_{\mathrm{FCI}}$. 
In other words, even if the SMA as an approximation to the magneto-roton mode in the FCI performs as good as in the FQH case, it is in general not possible to establish a correspondence between the magneto-roton and the SMA states.
%

\begin{figure}[t]
\includegraphics[width = 0.85\linewidth]{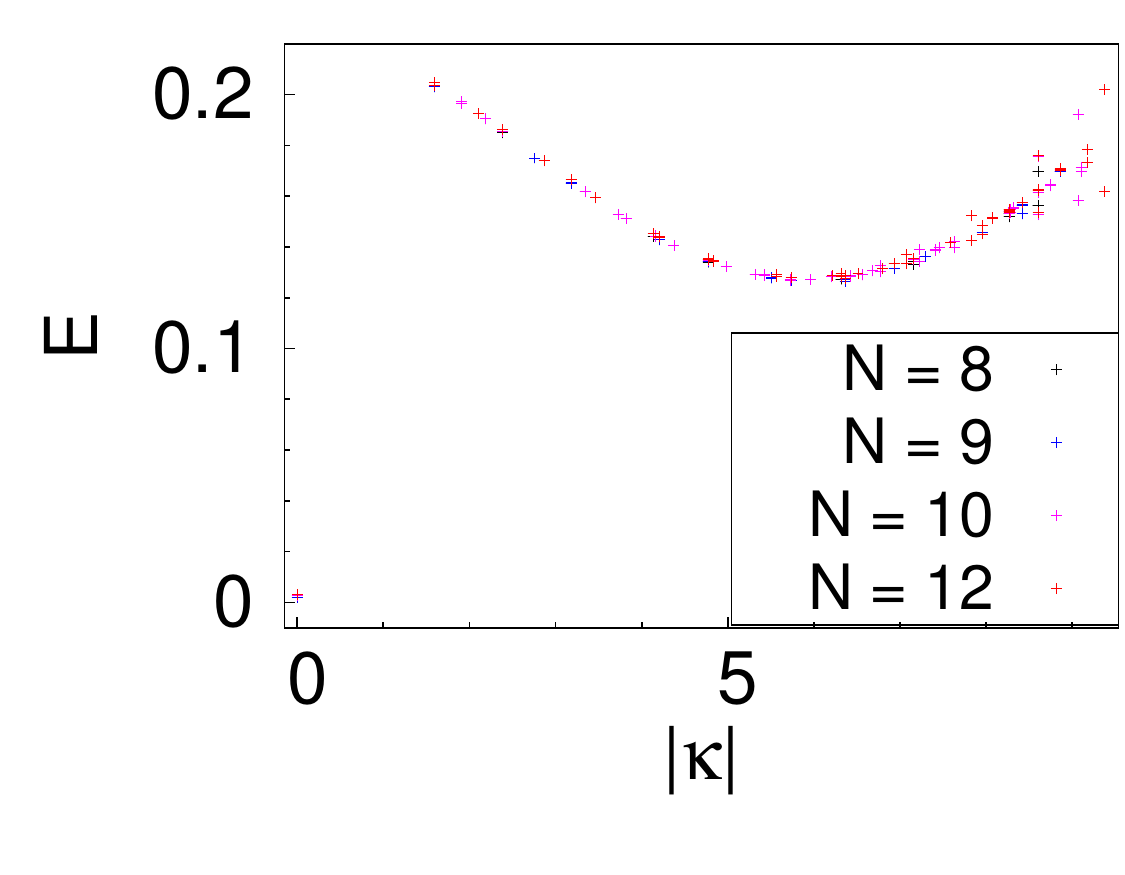}
\caption{Magneto-roton mode of the FCI system with up to $N = 12$ bosons, at half filling, obtained using the SMA as defined in Eq.~\pref{eq: criterion for FCI SMA}. We only show the states that obey the inequality defined in Eq.~\ref{eq: conduction with Kmax}.}
\label{fig:FCIEnergySMA}
\end{figure}

It is important to note that it is impossible to``unfold'' the eigenstates of the FCI to an enlarged Brillouin zone in any meaningful way, because of the lower translational symmetry of the FCI. It is thus impossible to reconstruct a nondegenerate magneto-roton dispersion (with one eigenstate per momentum quantum number) for the magneto-roton mode from the \emph{exact eigenstates} of the FCI.
In contrast, we should remember that the \emph{SMA states} of the FCI, as determined by the selection criterion~\eqref{eq: criterion for FCI SMA}, carry the reciprocal lattice vector $\bs{G}$ as an additional momentum quantum number. This additional information allows to unfold every SMA state into the respective Brillouin zone labelled by $\bs{G}$ (see Fig.~\ref{fig:FCIEnergySMA}) -- a procedure that could not be applied to the exact eigenstates.

We now turn to the interpretation of the cutoff $K_{\mathrm{max}}$ of Eq.~\eqref{eq: conduction with Kmax}. In general, the projected density operator $\rho_{\bs k}$ and its unitary counterpart $\rho_{\bs k}^{\mathrm{U}(1)}$ are not periodic under $\bs{k}\to\bs{k}+\bs{G}$, with $\bs{G}$ a reciprocal lattice vector. However, two SMA states generated with density operators at $\bs{k}$ and at $\bs{k}+\bs{G}$ are not orthogonal and can in fact have a large overlap. In the case of the ruby lattice model, and for all the system sizes that we have looked at, we find that any two states out of the set of SMA states that obey the constraint $|\bs{k}+\bs{G}- \bs{K_{\alpha}}| < 2\pi$ have mutual overlap smaller than $0.1$. Meanwhile, the overlap $\left|\braket{\Psi^{\mathrm{SMA}}_{\bs{k} + \bs{G},\alpha}}{\Psi^{\mathrm{SMA}}_{\bs{k} + \bs{G'},\alpha^\prime}}\right|^2$ between two SMA states that obey $|\bs{k}+\bs{G}- \bs{K_{\alpha}}| \geq 2\pi$ and $|\bs{k}+\bs{G}^\prime- \bs{K_{\alpha^\prime}}| < 2\pi$  is significantly larger. Note that the transition is rather abrupt, with these overlaps reaching $0.7$ or more already for $|\bs{k}+\bs{G}- \bs{K_{\alpha}}| = 2\pi$. If the magnetic translation symmetries were present, these overlaps would be $0$. Their large values thus reflect the absence of this symmetry in FCI, and we have to discard the corresponding states. This naturally sets the value of the cut-off parameter $K_{\mathrm{max}}$, that was introduced in Eq.~\eqref{eq: conduction with Kmax}, for the ruby lattice model to
\ba
K_{\mathrm{max}} = 2\pi,
\ea
as represented in Fig.~\ref{fig:FCISMARule}.

\begin{figure}[t]
\includegraphics[width = 0.8\linewidth]{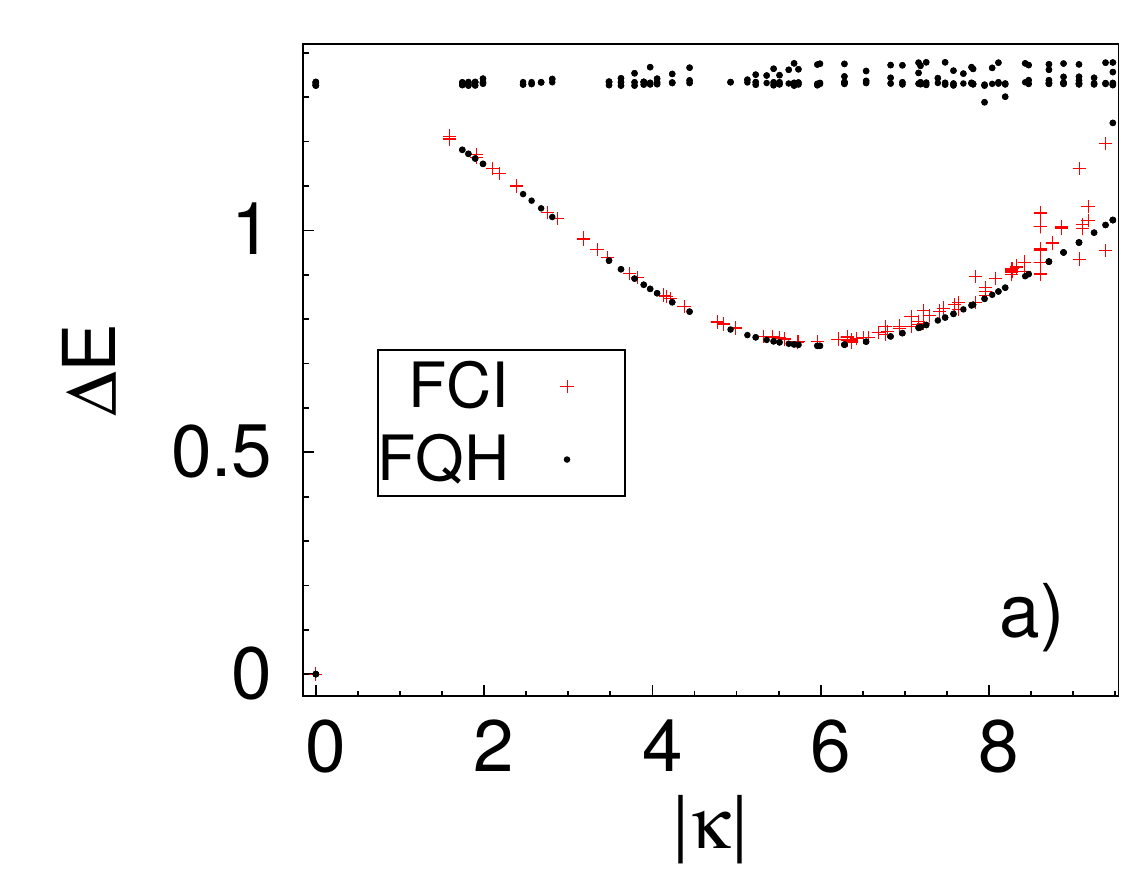}
\includegraphics[width = 0.8\linewidth]{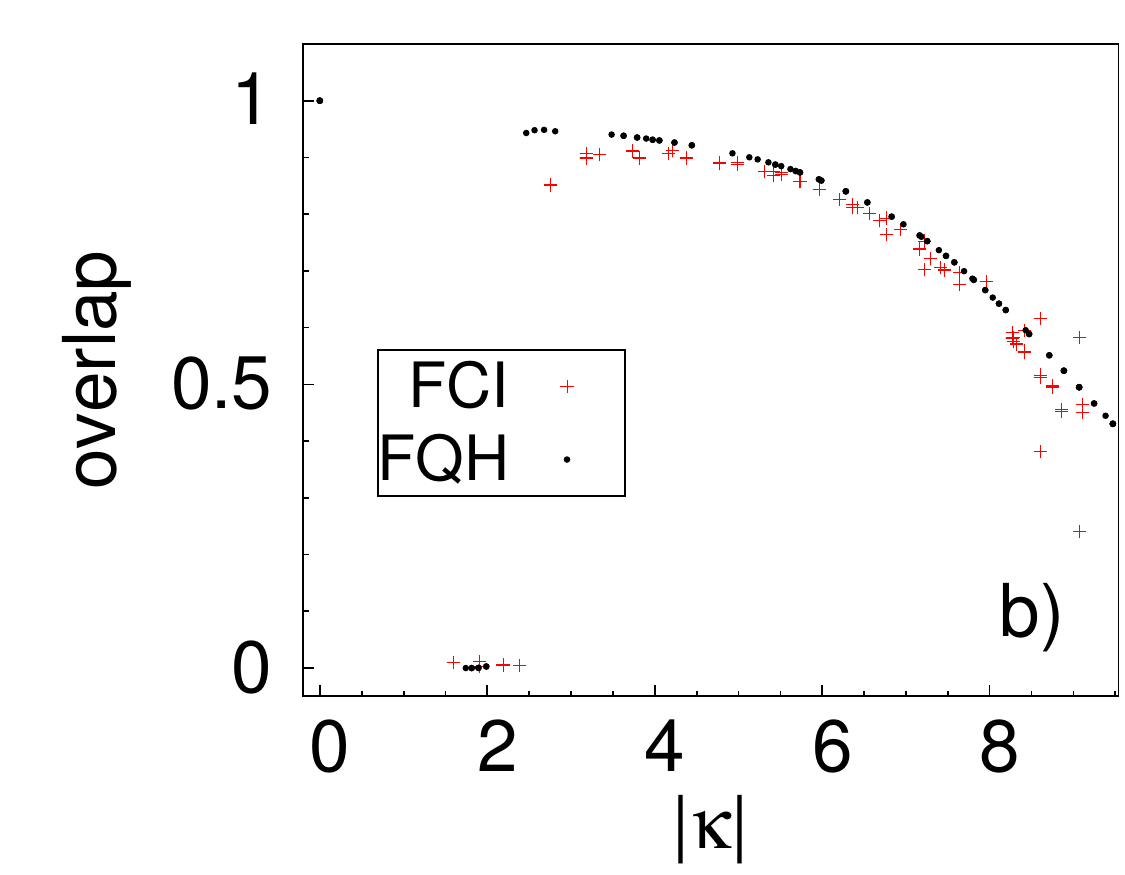}
\caption{
(a) Magneto-roton mode of the FQH and ruby lattice FCI systems at half filling, with respectively up to $N = 13$ (FQH) and $N = 12$ (FCI) particles, as computed using the SMA, for $|\bs{\kappa}| < \kappa_{\mathrm{max}} = \sqrt{\frac{2}{\sin \theta}}K_{\mathrm{max}}$. The FCI energies have been shifted by the ground state energy, and rescaled by the energy of the two-body problem (see Sec.~\ref{sec:Gap Extrapolation}). 
(b) Overlap of the SMA states of the FQH and FCI systems with the eigenstates obtained by exact diagonalization plotted as a function of $|\bs{\kappa}|$. In the FCI case, there may not be a one to one correspondence between the SMA states and the exact eigenstates. We therefore give the overlap with the whole FCI magneto-roton subspace, while the FQH overlaps are with individual states.}
\label{fig:FCIFQHEnergySMA}
\end{figure}

Having specified this set of rules, we are now equipped to answer the three benchmark questions (i)--(iii) for the SMA to the FCI magneto-roton mode.
We call $|\bs{\kappa}|$ the norm of the momentum vector $\bs{k}+\bs{G}- \bs{K_{\alpha}}$ up to the rescaling factor defined in Eq.~\eqref{eq: K-rescaling for data collapse}
\ba
|\bs{\kappa}| = \sqrt{\frac{2}{\sin \theta}} |\bs{k}+\bs{G}- \bs{K_{\alpha}}|
\ea
When the variational energy of the FCI SMA states selected by criterion~\eqref{eq: criterion for FCI SMA} are plotted as a function of $|\bs{\kappa}|$, one obtains an excellent agreement with the SMA dispersion of the FQH [see Fig.~\ref{fig:FCIFQHEnergySMA}~(a)]. Remarkably, the minimum of the FCI and FQH magneto-roton modes fall exactly at the same value of $|\bs{\kappa}|$. Similarly to the FQH case, only the SMA states with $|\bs{\kappa}| < 2\pi$ accurately approximate an exact eigenstate that belongs to the magneto-roton mode. Interestingly, almost no finite size effect is visible, even though FCIs are in general more susceptible to finite size effects than FQH systems. As pointed out in Sec.~\ref{sec:FCISMAConstruction}, imposing a cutoff $K_{\mathrm{max}}$ leads to generating less SMA modes than there are magneto-roton states. Fortunately, this does not reduce the number of magneto-roton states that are \emph{accurately} described by the SMA, as the cutoff lies at a larger value of $|\bs{\kappa}|$ than the minimum of the mode. We extract the energy minimum of the SMA mode, and compare it to the value obtained in Sec.~\ref{sec:Gap Extrapolation}. This variational value overestimates the value of the gap by about $20\%$, similar to the FQH SMA. For point (ii), Fig.~\ref{fig:FCIFQHEnergySMA}~(b) shows that the same separation in momenta $|\bs{k}+\bs{G}|$ also discriminates SMA states with a large and small overlap. We also note that the overlaps are slightly higher by about $1\%$ if the variant $\rho_{\bs{k}}^{\mathrm{U}(1)}$ of the density operator is used instead of $\rho_{\bs{k}}$. Moreover, the overlap values are not significantly smaller in the FCI case than their FQH counterparts. On average, the FCI overlaps are $5\%$ smaller than the FQH overlaps.

To address question (iii), we note that diagonalizing the Hamiltonian in the full bilinear subspace is not conceivable in the FCI case. Indeed, this method mixes large and small $|\bs{k}+\bs{G}|$ in the same momentum sectors. This leaves us with a spectrum that cannot be unfolded, and the variational states that give an acceptable approximation of the magneto-roton mode cannot be identified. However, one can compute the overlap of each exact magneto-roton eigenstate with the full bilinear subspace. Similarly to the FQH case, this overlap is only a few percents higher than that of the SMA states with the exact eigenstates, validating the SMA approach.

A crucial difference between the case of the FQH effect and the FCI that we would like to highlight, is that the density operator used to construct the SMA is uniquely defined for the former, while it contains the freedom to choose an embedding for the latter. The choice of embedding will in general influence the quality of the SMA. Particularly pathological are cases in which the orbital displacement vectors $\bs{r}_a$ are integer in units of the lattice constant. Then, the projected density operators share the periodicity of the reciprocal lattice in momentum space and will not suffice to build enough variational states for the SMA. To our knowledge, no model  for which such an embedding is natural hosts a robust Laughlin-like phase. The kagome lattice model has half integer $\bs{r}_a$ in units of the lattice constant, and thus presents some commensurability effect. However, even in this case, all SMA states within a circle of radius $K_{\mathrm{max}} = 2\pi$ are linearly independent. Unfortunately, as shown in Fig.~\ref{fig:FCISpectra8_9_10}, its magneto-roton mode is not as well defined as that of the ruby lattice system. Unsurprisingly, the eigenstates have a smaller overlap with the bilinear subspace (0.8 at best), and there is a lot of mixing between the states originating from different ground states. The variational SMA states, in turn, have a maximum overlap of $0.53$ with the exact eigenstates. Our efforts to tune the embedding away from the value given in Eq.~\eqref{eq: Kagome embedding} did not improve these overlaps significantly. Note that tuning the embedding for the ruby model would barely improve the overlaps in this case either. Indeed, they are already close to the maximal values that can be reached using all the bilinears originating from the same ground state.



\section{Conclusion}
\label{ref: Conclusion}

In summary, we have applied the SMA to FQH systems on the torus geometry, and shown how to select a reduced set of variational states to describe the magneto-roton mode.
We further identified a magneto-roton mode for FCI systems and developed the SMA for this case. We found that the FCI magneto-roton mode can be understood in close analogy to the FQH case, provided that the reduced translational symmetry of the FCI, as well as the freedom of embedding of the particle density in position space, are accurately accounted for. Remarkably, the SMA for FCIs provides an additional degree of freedom that allows to unfold the magneto-roton mode in an enlarged Brillouin zone, while the absence of magnetic translation symmetry in FCIs prevents any direct unfolding of the exact spectrum. This very important result credits the SMA with an additional purpose, on top of being a quantitatively accurate variational method. Interestingly, the dispersion relations of the FQH and FCI magneto-roton modes obtained using the SMA fall onto the same curve, and show almost no finite size effect.

Besides, we have given an extrapolation of the excitation gap of the $\nu=1/2$ bosonic Laughlin state for both the FQH and FCI cases. For systems of $10$ particles or more, the gap is almost independent of the system size. This robustness suggests that the extrapolation of the gap to the thermodynamic limit is very reliable. Moreover, the numerical value of the FCI gap falls within the range of uncertainty of the FQH gap, confirming the universal character of the gap of the Laughlin state in FCIs. 

\section*{Acknowledgement}
We are very grateful to B. A. Bernevig for illuminating discussions. We thank Y.-L. Wu for discussions. T.N. and C.R. acknowledge financial support from DARPA SPAWARSYSCEN Pacific N66001-11-1-4110. NR was supported by NSF CAREER DMR-095242, ONR-N00014-11-1-0635, MURI-130-6082, Packard Foundation, Keck grant and the Princeton Global Scholarship. N.R. and C.R. were supported by ANR-12-BS04-0002-02.

\bibliography{Magnetoroton}
\end{document}